\documentclass[11pt, a4paper]{article}
\usepackage[utf8]{inputenc}
\usepackage[T1]{fontenc}
\usepackage[margin=1in]{geometry}
\usepackage{amsmath,amssymb,amsfonts,mathtools}
\usepackage{graphicx}
\usepackage{booktabs}
\usepackage{array}
\usepackage[table]{xcolor}
\usepackage[numbers,sort&compress]{natbib}
\usepackage{subcaption}
\usepackage{hyperref}
\usepackage{authblk}
\usepackage{lmodern}
\hypersetup{colorlinks=true, allcolors=blue!60!black}

\usepackage{indentfirst}

\newcommand{\Bvec}{\vec{B}}
\newcommand{\Jvec}{\vec{J}}
\newcommand{\xvec}{\vec{x}}
\newcommand{\grad}{\nabla}
\newcommand{\curl}{\nabla \times}
\newcommand{\divg}{\nabla \cdot}

\newcommand{\tablestyle}{%
  \setlength{\heavyrulewidth}{0.8pt}%
  \setlength{\lightrulewidth}{0.5pt}%
  \setlength{\cmidrulewidth}{0.3pt}%
  \renewcommand{\arraystretch}{1.15}%
  \arrayrulecolor{black!35}}

\usepackage{pifont}
\newcommand{\cmark}{\ding{51}}
\newcommand{\xmark}{\ding{55}}

\title{\textbf{Non-axisymmetric MHD equilibria with an arbitrary number of field periods}}

\author[1,2]{Opal Issan}
\author[1]{Lucas Jos\'e}
\author[3]{Matt Landreman}
\author[1]{Jonathan Citrin}
\author[1]{Brendan Tracey}

\affil[1]{Google DeepMind, London, UK}
\affil[2]{Department of Mechanical and Aerospace Engineering, University of California San Diego, La Jolla, CA, USA}
\affil[3]{University of Maryland, College Park, MD, USA}

\date{\today}

\begin{document}
\maketitle

\begin{abstract}
Several closed-form non-axisymmetric magnetic fields with nested toroidal flux surfaces have recently been discovered as exact benchmarks for verifying three-dimensional magnetohydrodynamic equilibrium codes.
However, all of these solutions have at most two field periods ($N \le 2$), i.e., identical segments that repeat toroidally.
Here, we derive closed-form non-axisymmetric magnetohydrodynamic equilibria for an arbitrary integer number of field periods $N \geq 1$.
The equilibria have non-constant pressure and closed magnetic field lines with either zero rotational transform ($\iota = 0$) for $N \geq 1$ or integer rotational transform ($\iota = \pm N$) for $N \ge 3$.
For the reflection-symmetric $\iota = 0$ equilibria, we also construct a matched exterior vacuum field with continuous magnetic field across the plasma boundary, providing a benchmark for free-boundary equilibrium codes.
\end{abstract}

%===================================
\section{Introduction} \label{sec:intro}
%===================================
In Cartesian coordinates $\xvec \coloneq (x, y, z)$, the magnetic field $\Bvec(\xvec) \coloneq (B_x, B_y, B_z)$, current density $\Jvec(\xvec) \coloneq \curl\Bvec$, and scalar plasma pressure $p(\xvec)$ of a static ideal magnetohydrodynamic~(MHD) plasma satisfy
\begin{equation} \label{eq:mhd_system}
  \Jvec \times \Bvec = \grad p, \qquad \divg\Bvec = 0, \qquad \Bvec \cdot \grad\psi = 0,
\end{equation}
where the pressure $p(\psi)$ depends on the flux function $\psi(\xvec) \in [0, \delta]$ whose level sets form compact nested tori around a magnetic axis at $\psi = 0$ up to a boundary surface $\psi = \delta$.
We also use cylindrical coordinates $(R, \phi, z)$ defined by $x= R\cos(\phi)$ and $y= R\sin(\phi)$.

Assuming axisymmetry ($\partial_\phi = 0$), the static ideal MHD equations~\eqref{eq:mhd_system} simplify to the two-dimensional Grad--Shafranov equation, which admits many known exact solutions~\cite{Shafranov1966, Solovev1968, Zheng1996, CerfonFreidberg2010}.
Without axisymmetry, resonant Pfirsch--Schl\"uter currents on rational surfaces~\cite{Boozer1981, Helander2014} prompted Grad~\cite{Grad1967, Grad1985} to conjecture that smooth toroidal equilibria with non-constant pressure do not exist.
Nevertheless, Lortz~\cite{lortz1970existenz} proved more than five decades ago the existence of reflection-symmetric equilibria with zero rotational transform ($\iota = 0$) via a fixed-point iteration, where $\iota$ is the ratio of poloidal to toroidal turns of a magnetic field line (positive or negative depending on the winding direction).
More recently, \citet{GomezSerrano2026} established the existence of smooth non-axisymmetric toroidal equilibria with discrete rotational symmetry and infinite rotational transform ($\iota = \infty$).
However, neither proof provides explicit closed-form solutions.

Several closed-form three-dimensional toroidal magnetic fields with nested flux surfaces were discovered very recently~\cite{Landreman2026, Issan2026, SenguptaVacuum2026, Lopez2026}.
In the vacuum limit of $\Jvec = \vec{0}$ and $\grad p = \vec{0}$, \citet{SenguptaVacuum2026} constructed closed-form magnetic fields with constant (rational or irrational) or sheared rotational transform using Jacobi elliptic functions, and \citet{Lopez2026} constructed a closed-form magnetic field with $\iota = -1$ via a cubic Weierstrass polynomial.
For non-constant pressure, \citet{Landreman2026} derived two closed-form stellarator-symmetric MHD equilibria, one with $\iota = 2$ and one with sheared rotational transform~$\iota'(\psi) \neq 0$, which \citet{Issan2026} generalized to equilibria without stellarator symmetry~\cite{DewarHudson1998}.

A non-axisymmetric toroidal equilibrium has $N$ field periods if it satisfies the following discrete rotational symmetry
\begin{equation} \label{eq:nfp_symmetry}
  \psi\big(R, \phi + 2\pi/N, z\big) = \psi(R, \phi, z), \qquad
  B_j\big(R, \phi + 2\pi/N, z\big) = B_j(R, \phi, z) \quad \text{for } j \in \{R, \phi, z\}.
\end{equation}
To the best of our knowledge, all recent closed-form solutions~\cite{Landreman2026, Issan2026, SenguptaVacuum2026, Lopez2026} are restricted to at most two field periods ($N \le 2$).
In practice, most stellarators are designed with $N \ge 3$ field periods, including NCSX ($N = 3$)~\cite{Zarnstorff2001}, HSX and TJ-II ($N = 4$)~\cite{Anderson1995, Alejaldre1999}, Wendelstein~7-X ($N = 5$)~\cite{Grieger1992, Klinger2019}, and LHD ($N = 10$)~\cite{Iiyoshi1999}.
Extending analytical solutions to $N \ge 3$ therefore broadens the variety of exact benchmarks available for verifying three-dimensional equilibrium codes such as VMEC~\cite{HirshmanWhitson1983}, DESC~\cite{dudt2020desc}, GVEC~\cite{Hindenlang2025}, and SPEC~\cite{Hudson2012}.

In this paper, we construct closed-form non-axisymmetric MHD equilibria for an arbitrary integer number of field periods $N \geq 1$.
All solutions originate from a coordinate transformation that maps field-aligned coordinates to Cartesian coordinates via a three-harmonic Fourier expansion in the field-aligned toroidal angle.
This coordinate ansatz reduces the three-dimensional nonlinear MHD partial differential equations to a sequence of separable first-order linear ordinary differential equations and algebraic relations.
Solving this system of equations yields three distinct families of equilibria, each featuring nested toroidal flux surfaces, non-constant plasma pressure, stellarator symmetry, and closed magnetic field lines.
The first family has zero rotational transform ($\iota = 0$) for any number of field periods $N \geq 1$ and provides an explicit closed-form realization of the non-axisymmetric equilibria proven to exist by~\citet{lortz1970existenz}.
The second family features an $\iota = 0$ branch for $N \ge 2$ and an integer rotational transform $\iota = \pm N$ branch for $N \ge 3$, whereas the third family yields equilibria with $\iota = \pm N$ for $N \ge 3$.
The $\iota = \pm N$ branch of the second family is restricted to narrow cross-sections, while the third family has much broader ones with a larger plasma volume and beta.
Table~\ref{tab:comparison} compares the equilibria derived in this work with recent closed-form non-axisymmetric solutions~\cite{Landreman2026, Issan2026, SenguptaVacuum2026, Lopez2026}.

This paper is organized as follows.
Section~\ref{sec:equilibrium} presents the coordinate map ansatz and the three families of equilibria.
Section~\ref{sec:verification} verifies the nested toroidal flux surfaces, Hamada condition, and discrete symmetries.
Section~\ref{sec:geometry} characterizes the magnetic axis, enclosed plasma volume, volume-averaged plasma beta, and ideal-MHD interchange stability.
Section~\ref{sec:vacuum} constructs a matched exterior vacuum field for $\iota = 0$ equilibria, providing a benchmark for free-boundary equilibrium codes.
Section~\ref{sec:summary} summarizes the results and discusses future work, and Appendix~\ref{app:derivation} presents a detailed derivation of the equilibria.

\begin{table}[htbp]
  \centering
  \footnotesize
  \caption{Comparison of closed-form non-axisymmetric toroidal solutions.}
  \label{tab:comparison}
  \tablestyle
  \begin{tabular}{@{}>{\raggedright\arraybackslash}p{0.14\textwidth}>{\raggedright\arraybackslash}p{0.14\textwidth}>{\raggedright\arraybackslash}p{0.14\textwidth}>{\raggedright\arraybackslash}p{0.15\textwidth}>{\raggedright\arraybackslash}p{0.12\textwidth}>{\raggedright\arraybackslash}p{0.16\textwidth}@{}}
    \toprule
    \textbf{Property} & \textbf{Landreman} \cite{Landreman2026} & \textbf{Issan et al.} \cite{Issan2026} & \textbf{Sengupta et al.} \cite{SenguptaVacuum2026} & \textbf{L\'opez} \cite{Lopez2026} & \textbf{Present Work} \\
    \midrule
    Physical regime & $\grad p \neq \vec{0}$, $\Jvec \neq \vec{0}$ & $\grad p \neq \vec{0}$, $\Jvec \neq \vec{0}$ & $\grad p = \vec{0}$, $\Jvec = \vec{0}$ & $\grad p = \vec{0}$, $\Jvec = \vec{0}$ & $\grad p \neq \vec{0}$, $\Jvec \neq \vec{0}$ \\
    \cmidrule{1-6}
    Analytical form & Elementary & Elementary & Jacobi elliptic & Elementary & Elementary \\
    \cmidrule{1-6}
    Field periods~$N$ & $N = 2$ & $N = 2$ & $N = 2$ & $N = 1$ & $N \geq 1$ \\
    \cmidrule{1-6}
    Rotational transform $\iota(\psi)$ & $\iota = 2$ or sheared $\iota'(\psi) \neq 0$ & $\iota = 2$ or sheared $\iota'(\psi) \neq 0$ & Constant or sheared $\iota'(\psi) \neq 0$ & $\iota = -1$ & $\iota = 0$ ($N \ge 1$) or $\iota = \pm N$ ($N \ge 3$) \\
    \cmidrule{1-6}
    Stellarator symmetry & \cmark & \xmark/\cmark & \xmark/\cmark & \xmark & \cmark \\
    \bottomrule
  \end{tabular}
  \arrayrulecolor{black}
\end{table}

%=============================================================================
\section{Equilibria with an arbitrary number of field periods \texorpdfstring{$N$}{N}}  \label{sec:equilibrium}
%=============================================================================
We begin by introducing field-aligned curvilinear coordinates $(\rho, \zeta, \eta)$, consisting of a toroidal angle $\zeta \in [0, 2\pi)$ along the magnetic field lines together with radial and vertical field-line labels $\rho > 0$ and $\eta \in \mathbb{R}$, respectively.
The coordinate ansatz relating the field-aligned coordinates $(\rho, \zeta, \eta)$ to the Cartesian coordinates $(x, y, z)$ is
\begin{equation} \label{eq:xyz_embedding}
\begin{aligned}
    x + iy &= C_1(\rho)\,e^{i m_1\zeta} + \big(C_2(\rho) + i\eta\,C_4(\rho)\big)e^{i m_2\zeta} - C_3(\rho)\,e^{i m_3\zeta}, \\
    z &= \eta + C_5(\rho)\sin(N\zeta),
\end{aligned}
\end{equation}
where the toroidal mode numbers $(m_1, m_2, m_3)$ are
\begin{equation} \label{eq:mode_defs}
    m_1 \coloneq 1 + (k - 1)N, \qquad m_2 \coloneq 1 + k N, \qquad m_3 \coloneq 1 + (k + 1)N, \qquad \text{for } k \in \{1, 0, -1\}.
\end{equation}
For a length-scale parameter $\rho_0 > 0$ and magnetic axis location $(\rho, \eta) = (\rho_a, 0)$, the auxiliary functions in Eq.~\eqref{eq:xyz_embedding} are
\begin{equation} \label{eq:C}
\begin{aligned}
  C_1(\rho) &\coloneq \frac{\operatorname{sign}(m_1)\,\rho_0}{m_1^2}\left(\frac{\rho}{\rho_0}\right)^{m_1}, &\qquad
  C_2(\rho) &\coloneq C_2(\rho_a)\left(\frac{\rho_a}{\rho}\right)^{N}\frac{D_-(\rho_a)}{D_-(\rho)}, \\[1.2ex]
  C_3(\rho) &\coloneq \frac{\rho_0}{m_3^2}\left(\frac{\rho_0}{\rho}\right)^{m_3}, &\qquad
  C_4(\rho) &\coloneq \frac{N^2 C_5(\rho)}{D_+(\rho)}, \\[1.2ex]
  C_5(\rho) &\coloneq \frac{2\,\rho_0}{N\sqrt{|m_1|\,m_3}}\left(\frac{\rho_0}{\rho}\right)^{N}, & &
\end{aligned}
\end{equation}
where
\begin{equation} \label{eq:D_C2_defs}
\begin{aligned}
    D_\pm(\rho) &\coloneq m_1(m_1 + m_2)C_1(\rho) \pm m_3(m_3 + m_2)C_3(\rho),\\
    C_2(\rho_a) &\coloneq s\sqrt{\frac{\big(m_1 C_1(\rho_a) + m_3 C_3(\rho_a)\big)\big(m_1^2 C_1(\rho_a) - m_3^2 C_3(\rho_a)\big) D_-(\rho_a)}{m_2^3 D_+(\rho_a)}}, \qquad  s \in \{-1, 1\}.
\end{aligned}
\end{equation}
Aligning the magnetic field with the $\zeta$-coordinate curves via $\Bvec \coloneq \partial_\zeta\xvec$ results in
\begin{equation} \label{eq:B_cartesian_param}
\begin{aligned}
  B_x + iB_y &= i m_1 C_1(\rho)\,e^{i m_1\zeta} + i m_2\big(C_2(\rho) + i\eta\,C_4(\rho)\big)e^{i m_2\zeta} - i m_3 C_3(\rho)\,e^{i m_3\zeta}, \\
  B_z &= N C_5(\rho)\cos(N\zeta).
\end{aligned}
\end{equation}
The flux function $\psi(\rho, \eta)$ and plasma pressure $p(\psi)$ with on-axis pressure $p_0 > 0$ are
\begin{equation} \label{eq:psi_p}
\begin{aligned}
  \psi(\rho, \eta) &\coloneq \frac{H(\rho, \eta) - H(\rho_a, 0)}{2},  \\
   p(\psi) &\coloneq p_0 - 2\psi, \\
  H(\rho, \eta) &\coloneq \big(m_1 C_1(\rho) + m_3 C_3(\rho)\big)^2 + m_2^2\big(C_2(\rho)^2 + \eta^2 C_4(\rho)^2\big).
\end{aligned}
\end{equation}
Appendix~\ref{app:derivation} derives Eqs.~\eqref{eq:xyz_embedding}--\eqref{eq:psi_p} from the static ideal MHD equations~\eqref{eq:mhd_system}.
Briefly, the ansatz~\eqref{eq:xyz_embedding} arises from perturbing the axisymmetric map  of $x + iy = \rho e^{i\zeta}$ and $z = \eta$ with an $N$-periodic vertical oscillation $\sin(N\zeta)$ in $z$, which requires three harmonics $(m_1, m_2, m_3)$ spaced by $N$ in $x + iy$ to satisfy the force-balance and divergence-free equations~\eqref{eq:mhd_system}.

The coordinate Jacobian determinant $\mathcal{J}$ of Eq.~\eqref{eq:xyz_embedding} is independent of $\zeta$, i.e. 
\begin{equation} \label{eq:jacobian_exact}
\begin{aligned}
  \mathcal{J}(\rho, \eta) &\coloneq \det\big(\partial\xvec/\partial(\rho,\zeta,\eta)\big) \\
  &= \frac{D_-(\rho)}{\rho D_+(\rho)}\Big(\big(m_1 C_1(\rho) + m_3 C_3(\rho)\big)^2 - m_2^2 \eta^2 C_4(\rho)^2\Big) - \frac{D_+(\rho)}{\rho D_-(\rho)}m_2^2 C_2(\rho)^2,
\end{aligned}
\end{equation}
which ensures that $\divg\Bvec = \mathcal{J}^{-1}\partial_\zeta\mathcal{J} = 0$.
For the coordinate map in Eq.~\eqref{eq:xyz_embedding} to remain smooth and invertible on the toroidal domain, the denominators $D_\pm(\rho)$ in Eq.~\eqref{eq:D_C2_defs} and the Jacobian determinant $\mathcal{J}(\rho, \eta)$ in Eq.~\eqref{eq:jacobian_exact} must remain non-zero.

We categorize the equilibria into three families based on $k \in \{1, 0, -1\}$ in Eq.~\eqref{eq:mode_defs}.
In all numerical examples in this paper, we set $\rho_0 = 1$ and $p_0 = 2\delta$ for simplicity, so that the pressure vanishes on the boundary flux surface, $p(\delta) = 0$.
The remaining parameters are the number of field periods $N$, the sign $s \in \{-1, 1\}$ in Eq.~\eqref{eq:D_C2_defs}, the magnetic axis parameter $\rho_a > 0$, and the boundary flux $\delta > 0$, which are restricted by $D_\pm(\rho)$ and $\mathcal{J}(\rho, \eta)$ remaining non-zero.
Below, we state the conditions on $\rho_a$ for each family and summarize their properties in Table~\ref{tab:families}.

\begin{figure}[!htb]
  \centering
  \begin{subfigure}[t]{0.32\textwidth}
    \centering
    \caption{$N = 1$}
    \includegraphics[width=\textwidth]{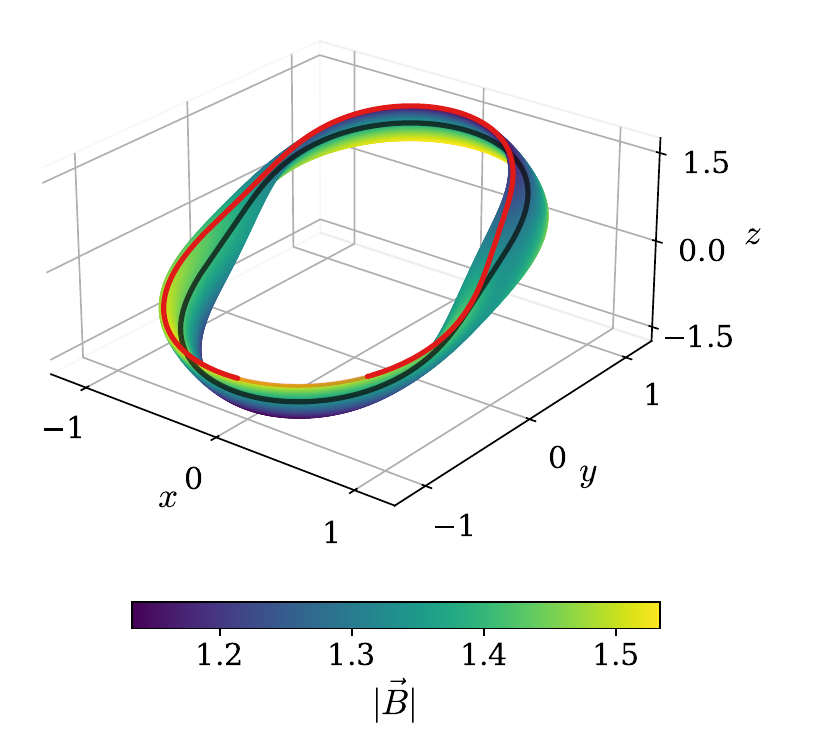}
    \label{fig:3d_n1}
  \end{subfigure}
  \begin{subfigure}[t]{0.32\textwidth}
    \centering
    \caption{$N = 2$}
    \includegraphics[width=\textwidth]{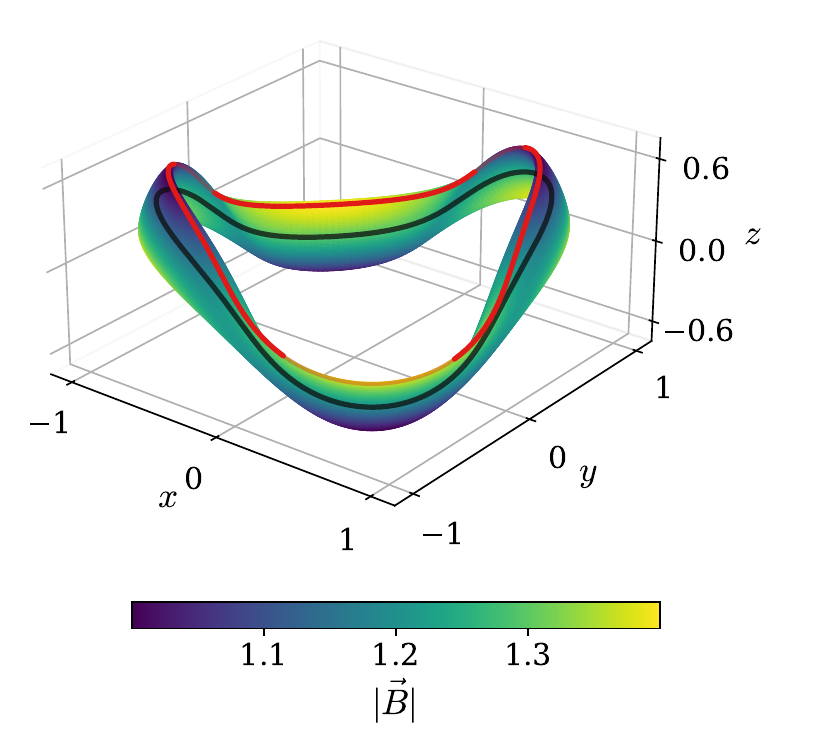}
    \label{fig:3d_n2}
  \end{subfigure}
  \begin{subfigure}[t]{0.32\textwidth}
    \centering
    \caption{$N = 3$}
    \includegraphics[width=\textwidth]{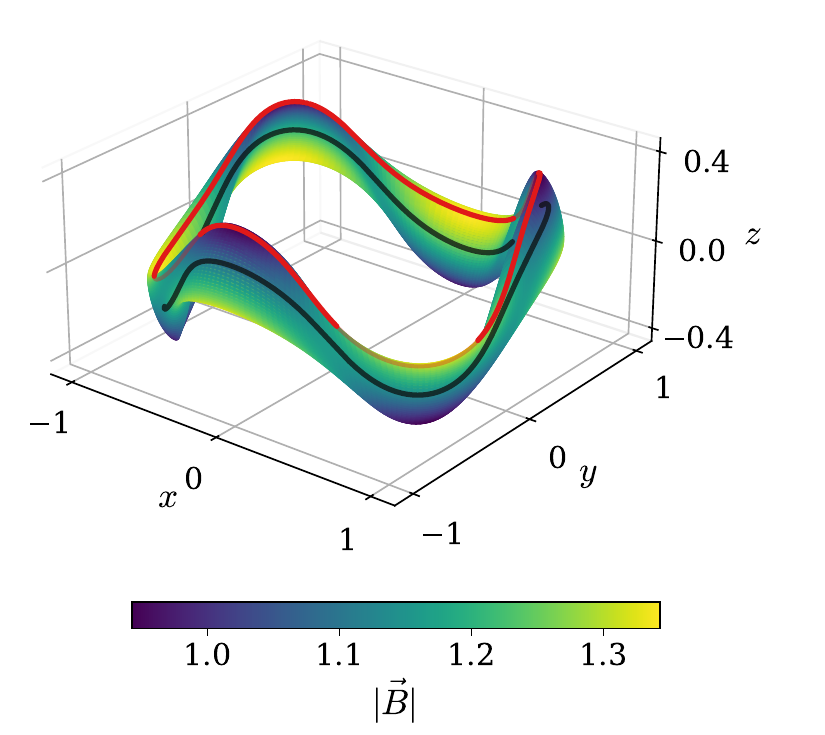}
    \label{fig:3d_n3}
  \end{subfigure}
      \\[0.6em]
  \begin{subfigure}[t]{0.32\textwidth}
    \centering
    \caption{$N = 4$}
    \includegraphics[width=\textwidth]{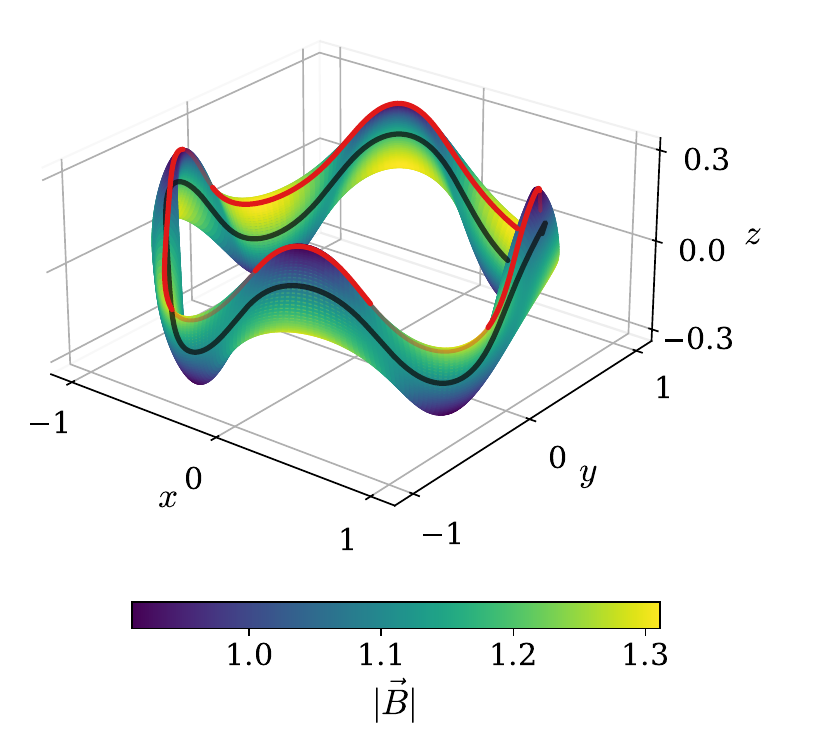}
    \label{fig:3d_n4}
  \end{subfigure}
  \begin{subfigure}[t]{0.32\textwidth}
    \centering
    \caption{$N = 5$}
    \includegraphics[width=\textwidth]{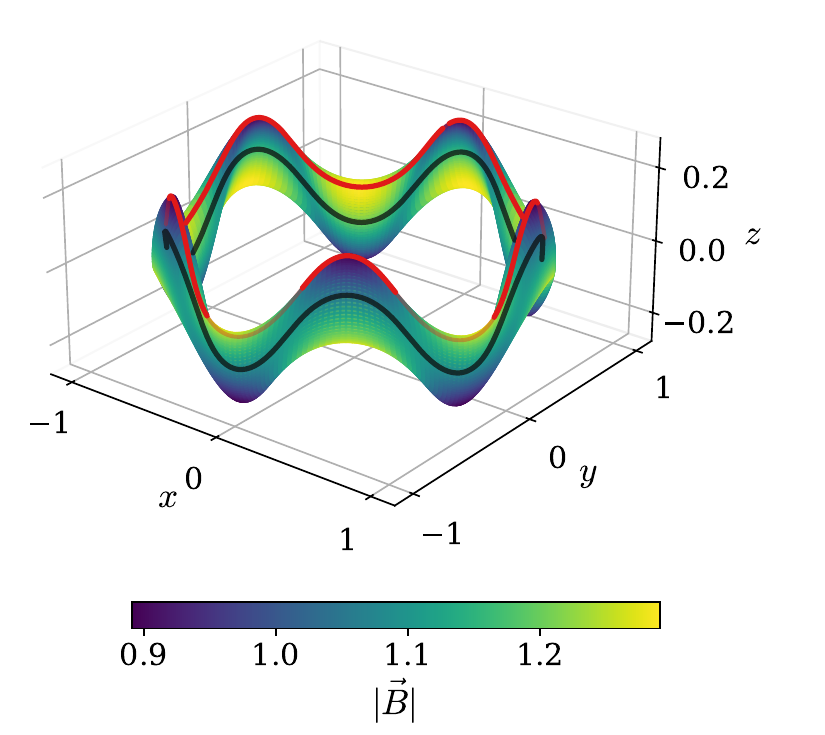}
    \label{fig:3d_n5}
  \end{subfigure}
  \begin{subfigure}[t]{0.32\textwidth}
    \centering
    \caption{$N = 6$}
    \includegraphics[width=\textwidth]{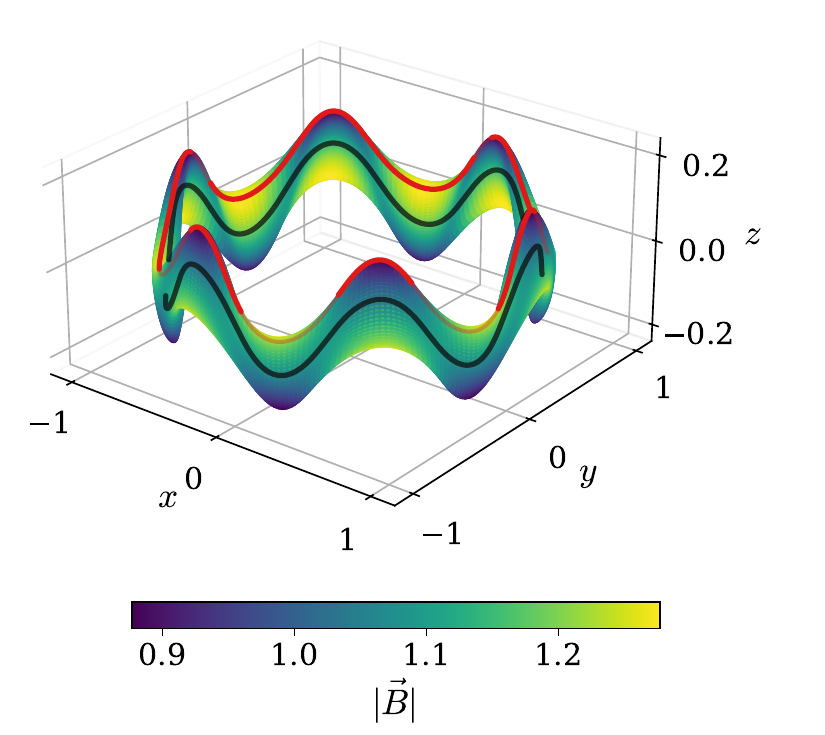}
    \label{fig:3d_n6}
  \end{subfigure}
  \caption{Family~1 outer toroidal flux surfaces $\psi = \delta = 0.02$ with $\rho_{a} = \rho_{0}$ illustrating the magnetic amplitude $|\Bvec|$ alongside the magnetic axis (black curve) and a magnetic field line on the boundary surface (red curve) for various numbers of field periods $N$.}
  \label{fig:3d_surfaces}
\end{figure}

\paragraph{Family 1.}
Setting $k=1$ results in $(m_1, m_2, m_3) = (1, N+1, 2N+1)$, and choosing $\rho_a \ge \rho_0$ yields nested toroidal flux surfaces with $\iota = 0$ for any number of field periods $N \geq 1$.
Figure~\ref{fig:3d_surfaces} shows the boundary flux surface $\psi = \delta = 0.02$ shaded by the magnetic field amplitude $|\Bvec|$ along with the magnetic axis (black curve) and a magnetic field line on the boundary surface (red curve) for various numbers of field periods $N$.

\paragraph{Family 2.}
Setting $k = 0$ results in $(m_1, m_2, m_3) = (1-N, 1, 1+N)$ and requires $N \ge 2$.
For $N = 2$, any $\rho_a > 0$ yields nested toroidal flux surfaces with $\iota = 0$.
For $N \ge 3$, choosing $0 < \rho_a < \rho_{\mathrm{mid}}$ gives $\iota = 0$, whereas choosing $\rho_a > \rho_{\mathrm{mid}}$ with $D_+(\rho_a) > 0$ and $\partial_\rho^2 H(\rho_a, 0) > 0$ gives $\iota = -s N = \pm N$, where
\begin{equation*}
  \rho_{\mathrm{mid}} \coloneq \rho_0\left(\frac{(N + 2)(N - 1)^2}{(N - 2)(N + 1)^2}\right)^{1/4}.
\end{equation*}

\paragraph{Family 3.}
For $k = -1$ and $N \ge 3$, the mode numbers are $(m_1, m_2, m_3) = (1-2N, 1-N, 1)$, and choosing
\begin{equation*}
  0 < \rho_a < \rho_0\left(\frac{3N - 2}{(2N - 1)(N - 2)}\right)^{1/(2N - 2)} 
\end{equation*}
yields nested toroidal flux surfaces with $\iota = -s N = \pm N$ over a much wider range of $\delta$ than the $\iota = \pm N$ branch of Family~2.

\begin{figure}[!htb]
  \centering
  \begin{subfigure}[t]{0.38\textwidth}
    \centering
    \caption{Family~2: $N = 3$}
    \includegraphics[width=\textwidth]{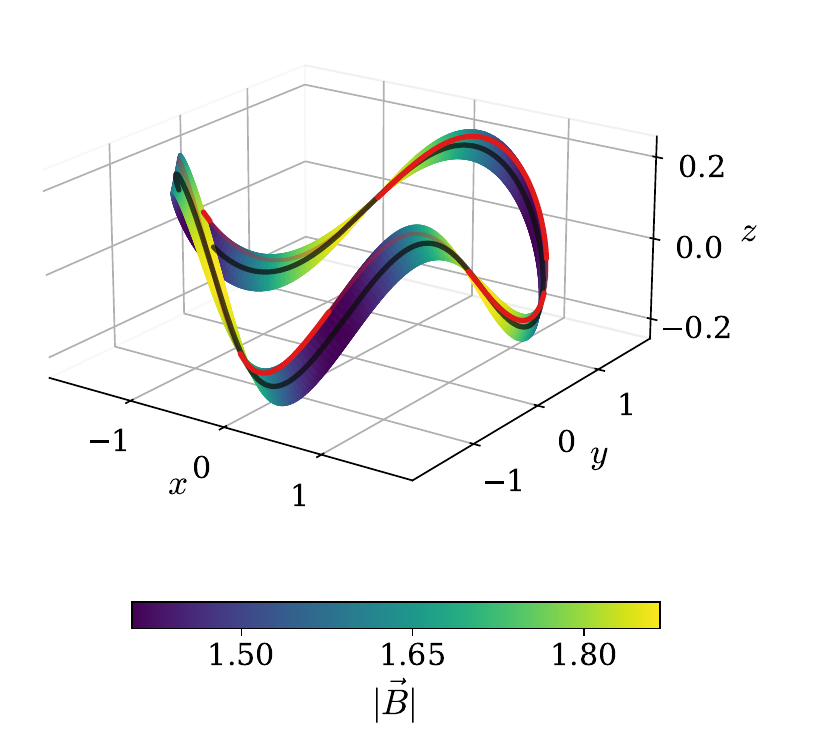}
    \label{fig:3d_fam2_pos_n3}
  \end{subfigure}
    \hspace{0.04\textwidth}
  \begin{subfigure}[t]{0.38\textwidth}
    \centering
    \caption{Family~2: $N = 5$}
    \includegraphics[width=\textwidth]{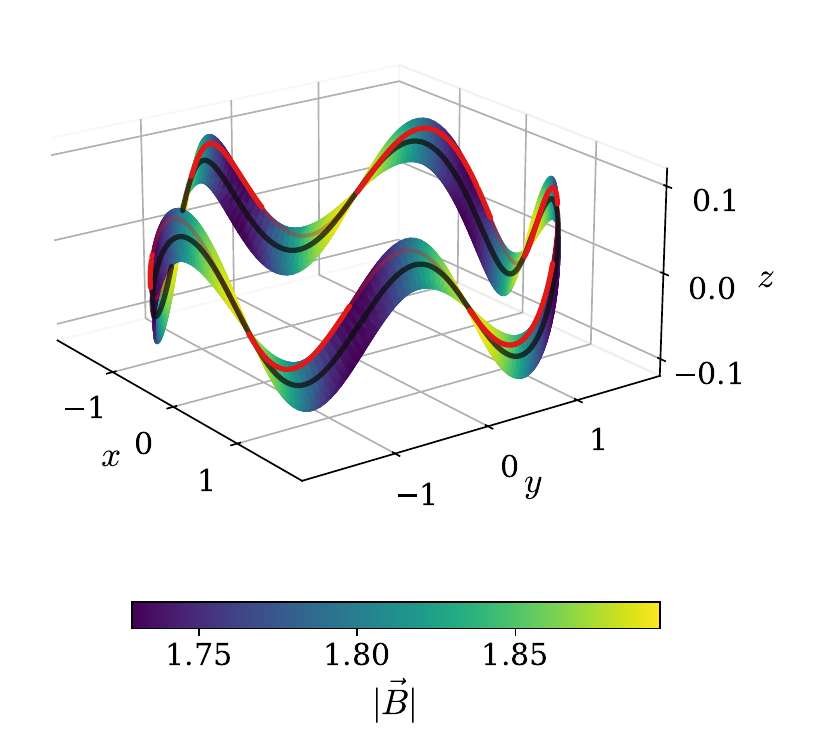}
    \label{fig:3d_fam2_pos_n5}
  \end{subfigure}
  \\[0.6em]
  \begin{subfigure}[t]{0.38\textwidth}
    \centering
    \caption{Family~3: $N = 3$}
    \includegraphics[width=\textwidth]{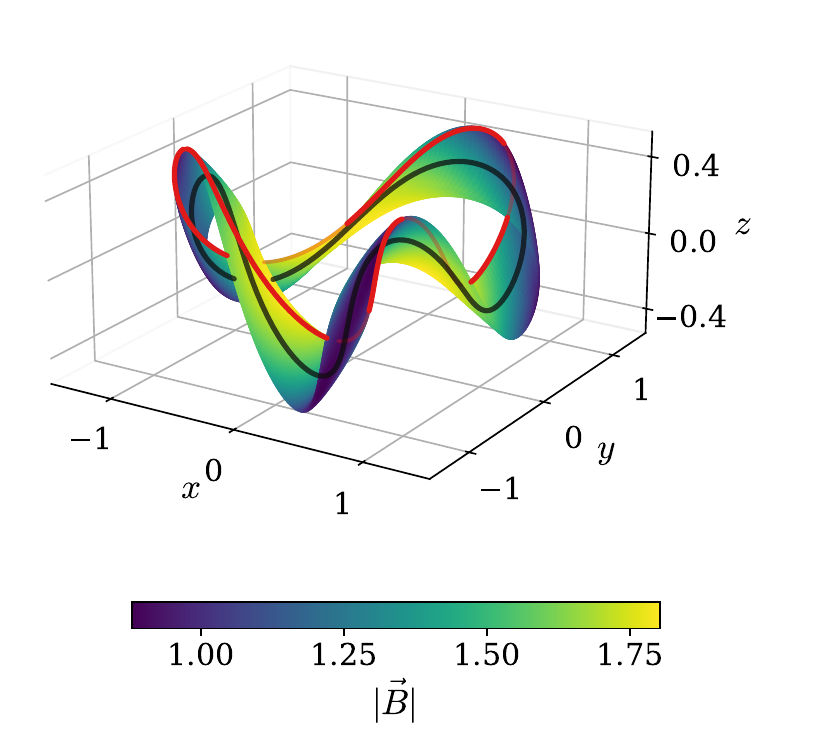}
    \label{fig:3d_fam3_pos_n3}
  \end{subfigure}
    \hspace{0.04\textwidth}
  \begin{subfigure}[t]{0.38\textwidth}
    \centering
    \caption{Family~3: $N = 5$}
    \includegraphics[width=\textwidth]{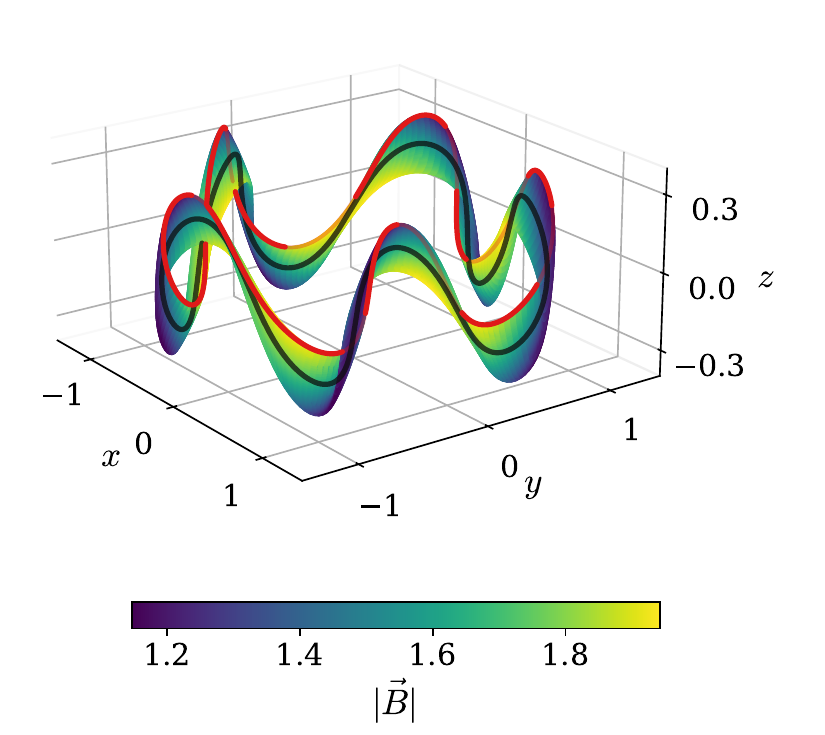}
    \label{fig:3d_fam3_pos_n5}
  \end{subfigure}
    \caption{Same as Figure~\ref{fig:3d_surfaces} with $\iota = N$ for Family~2 with $\delta = 0.005$ at (a)~$N = 3$ ($\rho_a = 1.12$) and (b)~$N = 5$ ($\rho_a = 1.04$) and for Family~3 with $\delta = 0.05$ at (c)~$N = 3$ ($\rho_a = 0.97$) and (d)~$N = 5$ ($\rho_a = 0.88$).}
  \label{fig:fam23_3d_surfaces}
\end{figure}

Figure~\ref{fig:fam23_3d_surfaces} shows the boundary flux surfaces $\psi = \delta$ shaded by $|\Bvec|$ together with the magnetic axis (black curve) and a boundary field line (red curve) for Family~2 ($\delta = 0.005$) and Family~3 ($\delta = 0.05$) at $N = \iota = 3$ and $N = \iota = 5$.
Although both equilibria have $\iota = N$, they represent distinct equilibria due to their different frequencies $(m_1, m_2, m_3)$ in Eq.~\eqref{eq:mode_defs}, with highest harmonic magnitude $N + 1$ in Family~2 versus $2N - 1$ in Family~3.
In addition, nested flux surfaces break down at a much smaller boundary flux in Family~2 ($\delta \approx 0.006$ at $N = 5$) than in Family~3 ($\delta \approx 0.05$ at $N = 5$).

\begin{table}[htbp]
  \centering
  \footnotesize
  \caption{Summary of the three families of equilibria generated by Eqs.~\eqref{eq:xyz_embedding}--\eqref{eq:psi_p}.}
  \label{tab:families}
  \tablestyle
  \begin{tabular}{@{}>{\raggedright\arraybackslash}p{0.25\textwidth}>{\raggedright\arraybackslash}p{0.21\textwidth}>{\raggedright\arraybackslash}p{0.24\textwidth}>{\raggedright\arraybackslash}p{0.22\textwidth}@{}}
    \toprule
    \textbf{Property} & \textbf{Family 1} & \textbf{Family 2} & \textbf{Family 3} \\
    \midrule
    Harmonic shift index $k$ & $k = 1$ & $k = 0$ & $k = -1$ \\
    \cmidrule{1-4}
    Mode numbers $(m_1,\; m_2,\; m_3)$ & $(1,\; N+1,\; 2N+1)$ & $(1-N,\; 1,\; 1+N)$ & $(1-2N,\; 1-N,\; 1)$ \\
    \cmidrule{1-4}
    Field periods $N$ & $N \ge 1$ & $N \ge 2$ & $N \ge 3$ \\
    \cmidrule{1-4}
    Rotational transform $\iota$ & $\iota = 0$ & $\iota = 0$ ($N \ge 2$),\newline $\iota = \pm N$ ($N \ge 3$) & $\iota = \pm N$ \\
    \bottomrule
  \end{tabular}
  \arrayrulecolor{black}
\end{table}

%=============================================================================
\section{Equilibria symmetries and verification} \label{sec:verification}
%=============================================================================
We first verify that the equilibria defined in Eqs.~\eqref{eq:xyz_embedding}--\eqref{eq:psi_p} form smooth nested toroidal flux surfaces in Section~\ref{subsec:flux_surfaces}, and then verify the Hamada condition in Section~\ref{subsec:hamada}.
We then demonstrate the discrete $N$-fold rotational symmetry in Section~\ref{subsec:nfp_symmetry}, reflection symmetry in Section~\ref{subsec:reflection_symmetry}, and stellarator symmetry in Section~\ref{subsec:stellarator_symmetry}.

%======================================================================
\subsection{Nested toroidal flux surfaces} \label{subsec:flux_surfaces}
%======================================================================
To verify that the level sets $0 < \psi \leq \delta$ form nested toroidal flux surfaces around the magnetic axis $(\rho, \eta) = (\rho_a, 0)$ in the parameter regimes of Section~\ref{sec:equilibrium}, we introduce the auxiliary coordinate
\begin{equation*}
  r(\rho) \coloneq \operatorname{sign}(\rho - \rho_a)\sqrt{H(\rho, 0) - H(\rho_a, 0)}.
\end{equation*}
The flux function $\psi$ in Eq.~\eqref{eq:psi_p} expressed in terms of $r(\rho)$ can be written as
\begin{equation*}
  \psi(\rho, \eta) = \frac{r(\rho)^2 + m_2^2 \eta^2 C_4(\rho)^2}{2}.
\end{equation*}
Then, introducing a field-line angle $\theta \in [0, 2\pi)$ via
\begin{equation*} 
  r(\rho) = \sqrt{2\psi}\cos(\theta), \qquad m_2\,\eta\,C_4(\rho) = \sqrt{2\psi}\sin(\theta),
\end{equation*}
maps each surface of constant $\psi$ to a circle of radius $\sqrt{2\psi}$ in the $(r, m_2\eta C_4)$ plane, so the flux surfaces are nested wherever the coordinate map in Eq.~\eqref{eq:xyz_embedding} remains smooth and invertible.

%======================================================================
\subsection{Hamada condition} \label{subsec:hamada}
%======================================================================
When magnetic field lines are closed and the pressure is non-constant, keeping the parallel current single-valued requires $\oint |\Bvec|^{-1}\,\mathrm{d}\ell$ to be constant on each flux surface, known as the Hamada condition~\cite{Newcomb1959, Hamada1962, lortz1970existenz}.
Because Eq.~\eqref{eq:xyz_embedding} traces each closed field line once over $\zeta \in [0, 2\pi)$ with arc-length element $\mathrm{d}\ell = |\partial\xvec/\partial\zeta|\,\mathrm{d}\zeta = |\Bvec|\,\mathrm{d}\zeta$, all three families of equilibria satisfy
\begin{equation*}
  \oint \frac{\mathrm{d}\ell}{|\Bvec|} = \int_0^{2\pi} \mathrm{d}\zeta = 2\pi,
\end{equation*}
which is constant across all field lines.

%======================================================================
\subsection{Discrete \texorpdfstring{$N$}{N}-fold rotational symmetry} \label{subsec:nfp_symmetry}
%======================================================================
An equilibrium has discrete $N$-fold rotational symmetry if its flux surfaces and magnetic field are invariant under a toroidal rotation by one field period, i.e. $(R, \phi, z) \mapsto (R, \phi + 2\pi/N, z)$, as defined in Eq.~\eqref{eq:nfp_symmetry}.
To verify this symmetry in field-aligned coordinates, we shift the toroidal angle by one field period $(\rho, \zeta, \eta) \mapsto (\rho, \zeta + 2\pi/N, \eta)$.
From Eq.~\eqref{eq:mode_defs}, each mode number $m_j$ for $j \in \{1, 2, 3\}$ equals $1$ plus an integer multiple of $N$, so
\begin{equation*}
  e^{i m_j(\zeta + 2\pi/N)} = e^{i 2\pi/N} e^{i m_j\zeta}.
\end{equation*}
Substituting $(\rho, \zeta + 2\pi/N, \eta)$ into Eq.~\eqref{eq:xyz_embedding} therefore rotates $(x, y)$ by $2\pi/N$ and leaves $z$ unchanged:
\begin{equation} \label{eq:discrete_rotation_xyz}
\begin{aligned}
  x(\rho, \zeta + 2\pi/N, \eta) &= R\cos(\phi + 2\pi/N), \\
  y(\rho, \zeta + 2\pi/N, \eta) &= R\sin(\phi + 2\pi/N), \\
  z(\rho, \zeta + 2\pi/N, \eta) &= z(\rho, \zeta, \eta),
\end{aligned}
\end{equation}
which confirms that $(\rho, \zeta, \eta) \mapsto (\rho, \zeta + 2\pi/N, \eta)$ maps $(R, \phi, z) \mapsto (R, \phi + 2\pi/N, z)$.
Since $\Bvec = \partial\xvec/\partial\zeta$, differentiating Eq.~\eqref{eq:discrete_rotation_xyz} with respect to $\zeta$ rotates $(B_x, B_y)$ by the same angle $2\pi/N$ and leaves $B_z$ unchanged, so the cylindrical components $B_R$, $B_\phi$, and $B_z$ are invariant under $(R, \phi, z) \mapsto (R, \phi + 2\pi/N, z)$.
Together with the $\zeta$-independence of $\psi(\rho, \eta)$, this confirms Eq.~\eqref{eq:nfp_symmetry}.

%======================================================================
\subsection{Reflection symmetry} \label{subsec:reflection_symmetry}
%======================================================================
In proving the existence of non-axisymmetric $\iota = 0$ equilibria, Lortz~\cite{lortz1970existenz} considered configurations that are symmetric across a poloidal plane $\phi = \phi_*$. 
Under the reflection $(R, \phi, z) \mapsto (R, 2\phi_* - \phi, z)$, reflection symmetry requires
\begin{equation} \label{eq:reflection_def}
\begin{aligned}
  \psi(R, 2\phi_* - \phi, z) &= \psi(R, \phi, z), &
  B_R(R, 2\phi_* - \phi, z) &= -B_R(R, \phi, z), \\
  B_\phi(R, 2\phi_* - \phi, z) &= B_\phi(R, \phi, z), &
  B_z(R, 2\phi_* - \phi, z) &= -B_z(R, \phi, z).
\end{aligned}
\end{equation}
Setting $\phi = \phi_*$ in Eq.~\eqref{eq:reflection_def} gives $B_R = B_z = 0$, so magnetic field lines intersect the plane $\phi = \phi_*$ perpendicularly.

In Family~1 with $\rho_a = \rho_0$, we have $C_2(\rho) = 0$ from Eq.~\eqref{eq:D_C2_defs}, so substituting $(\rho, \zeta, \eta) \mapsto (\rho, \pi/N - \zeta, \eta)$ into Eq.~\eqref{eq:xyz_embedding} results in
\begin{equation} \label{eq:reflection_symmetry_xyz}
\begin{aligned}
  x(\rho, \pi/N - \zeta, \eta) &= R\cos(\pi/N - \phi), \\
  y(\rho, \pi/N - \zeta, \eta) &= R\sin(\pi/N - \phi), \\
  z(\rho, \pi/N - \zeta, \eta) &= z(\rho, \zeta, \eta),
\end{aligned}
\end{equation}
which confirms that $(\rho, \zeta, \eta) \mapsto (\rho, \pi/N - \zeta, \eta)$ maps $(R, \phi, z) \mapsto (R, \pi/N - \phi, z)$.
Since $\Bvec = \partial\xvec/\partial\zeta$, differentiating Eq.~\eqref{eq:reflection_symmetry_xyz} with respect to $\zeta$ and projecting onto the cylindrical basis, together with the $\zeta$-independence of $\psi(\rho, \eta)$, confirms Eq.~\eqref{eq:reflection_def} across $\phi_* = \pi/2N$.

Family~1 with $\rho_a = \rho_0$ therefore provides an explicit closed-form realization of the reflection-symmetric $\iota = 0$ equilibrium class of~\citet{lortz1970existenz}.
By contrast, Family~1 with $\rho_a > \rho_0$, Family~2, and Family~3 do not satisfy reflection symmetry~\eqref{eq:reflection_def} across any poloidal plane.

%======================================================================
\subsection{Stellarator symmetry} \label{subsec:stellarator_symmetry}
%======================================================================
An equilibrium has stellarator symmetry~\cite{DewarHudson1998} if it is invariant under a $\pi$-rotation about the $x$-axis ($\phi = 0$). 
This rotation maps $(R, \phi, z) \mapsto (R, -\phi, -z)$ and requires
\begin{equation} \label{eq:stellarator_def}
\begin{aligned}
  \psi(R, -\phi, -z) &= \psi(R, \phi, z), &
  B_R(R, -\phi, -z) &= -B_R(R, \phi, z), \\
  B_\phi(R, -\phi, -z) &= B_\phi(R, \phi, z), &
  B_z(R, -\phi, -z) &= B_z(R, \phi, z).
\end{aligned}
\end{equation}
To verify this symmetry in field-aligned coordinates for all three families of equilibria, we substitute $(\rho, \zeta, \eta) \mapsto (\rho, -\zeta, -\eta)$ into Eq.~\eqref{eq:xyz_embedding}, i.e.
\begin{equation} \label{eq:stellarator_symmetry_xyz}
\begin{aligned}
  x(\rho, -\zeta, -\eta) &= R\cos(-\phi), \\
  y(\rho, -\zeta, -\eta) &= R\sin(-\phi), \\
  z(\rho, -\zeta, -\eta) &= -z(\rho, \zeta, \eta),
\end{aligned}
\end{equation}
which confirms that $(\rho, \zeta, \eta) \mapsto (\rho, -\zeta, -\eta)$ maps $(R, \phi, z) \mapsto (R, -\phi, -z)$.
Since $\Bvec = \partial\xvec/\partial\zeta$, differentiating Eq.~\eqref{eq:stellarator_symmetry_xyz} with respect to $\zeta$ gives
\begin{equation*}
  \Bvec(\rho, -\zeta, -\eta) = \big({-B_x(\rho, \zeta, \eta)},\, B_y(\rho, \zeta, \eta),\, B_z(\rho, \zeta, \eta)\big).
\end{equation*}
Projecting $\Bvec$ onto the cylindrical basis, together with $\psi(\rho, -\eta) = \psi(\rho, \eta)$, confirms Eq.~\eqref{eq:stellarator_def}.

\begin{figure}[htb!]
  \centering
  \begin{subfigure}[t]{0.42\textwidth}
    \centering
    \caption{Family~1: $N = 3$ ($\iota = 0$)}
    \includegraphics[width=\textwidth]{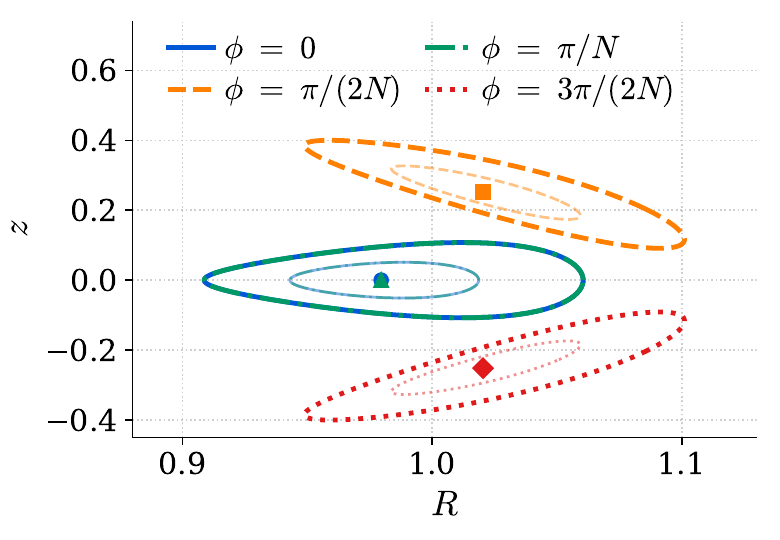}
    \label{fig:rz_fam1_zero_n3}
  \end{subfigure}
  \hspace{0.04\textwidth}
  \begin{subfigure}[t]{0.42\textwidth}
    \centering
    \caption{Family~2: $N = 3$ ($\iota = 0$)}
    \includegraphics[width=\textwidth]{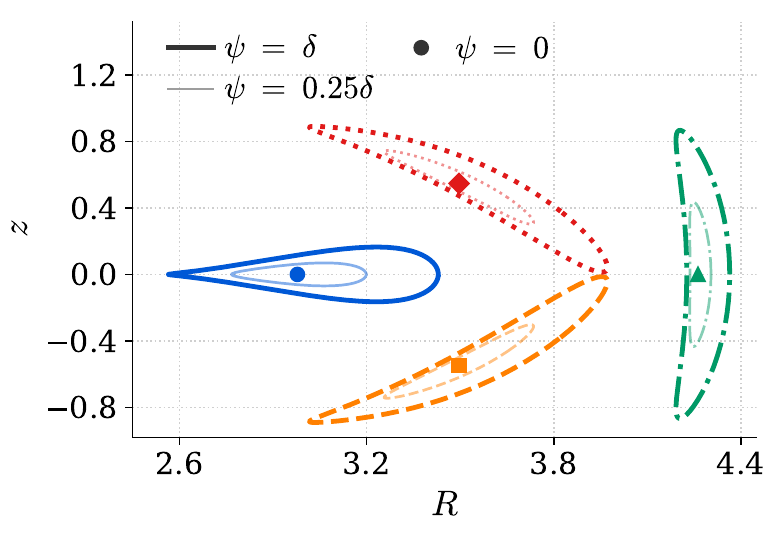}
    \label{fig:rz_fam2_zero_n3}
  \end{subfigure}
  \\[0.6em]
  \begin{subfigure}[t]{0.42\textwidth}
    \centering
    \caption{Family~2: $N = 3$ ($\iota = 3$)}
    \includegraphics[width=\textwidth]{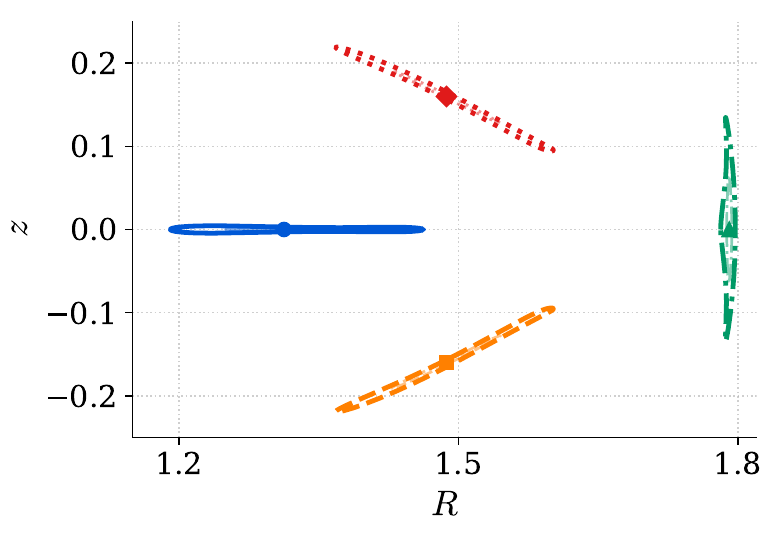}
    \label{fig:rz_fam2_pos_n3}
  \end{subfigure}
  \hspace{0.04\textwidth}
  \begin{subfigure}[t]{0.42\textwidth}
    \centering
    \caption{Family~3: $N = 3$ ($\iota = 3$)}
    \includegraphics[width=\textwidth]{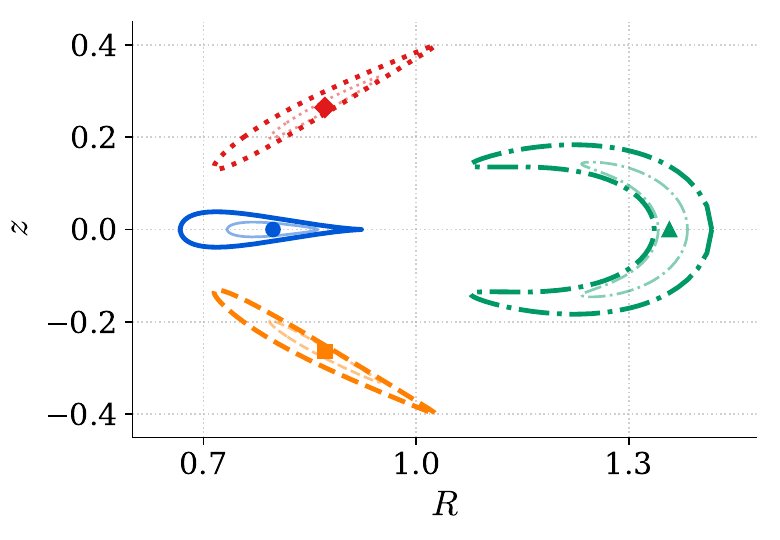}
    \label{fig:rz_fam3_pos_n3}
  \end{subfigure}
  \caption{Poloidal cross-sections at $N = 3$ for (a)~Family~1 ($\rho_a = 1$, $\delta = 0.02$, $\iota = 0$), (b)~Family~2 ($\rho_a = 0.75$, $\delta = 0.32$, $\iota = 0$), (c)~Family~2 ($\rho_a = 1.12$, $\delta = 0.02$, $\iota = 3$), and (d)~Family~3 ($\rho_a = 0.97$, $\delta = 0.05$, $\iota = 3$).}
  \label{fig:poloidal_rz}
\end{figure}

Figure~\ref{fig:poloidal_rz} shows the poloidal cross-sections of the flux surfaces across four cylindrical azimuthal slices $\phi \in \{0, \pi/2N, \pi/N, 3\pi/2N\}$ at $N = 3$ across all three families of equilibria.
By stellarator symmetry~\eqref{eq:stellarator_def} and discrete $N$-fold rotational symmetry~\eqref{eq:nfp_symmetry}, the cross-sections at $\phi = 0$ and $\phi = \pi/N$ are each symmetric across $z = 0$, while the slices at $\phi = \pi/2N$ and $\phi = 3\pi/2N$ are vertical reflections of one another across $z = 0$.
In Figure~\ref{fig:rz_fam1_zero_n3} ($C_2(\rho) = 0$), reflection symmetry~\eqref{eq:reflection_def} across $\phi_* = \pi/2N$ makes the $\phi = 0$ and $\phi = \pi/N$ cross-sections coincide, whereas in Figures~\ref{fig:rz_fam2_zero_n3}--\ref{fig:rz_fam3_pos_n3} ($C_2(\rho_a) \neq 0$) the $\phi = 0$ and $\phi = \pi/N$ cross-sections are distinct.
In Figure~\ref{fig:rz_fam2_pos_n3}, the cross-sections on the $\iota = N$ branch of Family~2 are much narrower because $|\mathcal{J}|$ is small near the magnetic axis, which compresses the flux surfaces and restricts $\delta$ to small values where the flux surfaces remain nested.

%=============================================================================
\section{Equilibria geometry and physical properties} \label{sec:geometry}
%=============================================================================
We now characterize the magnetic axis geometry in Section~\ref{subsec:axis}, the enclosed plasma volume $V(\delta)$ and volume-averaged plasma beta $\beta_V(\delta)$ in Section~\ref{subsec:vol_beta}, and ideal-MHD interchange stability in Section~\ref{subsec:stability}.

%======================================================================
\subsection{Magnetic axis}\label{subsec:axis}
\begin{figure}[htbp]
  \centering
  \begin{subfigure}[t]{0.32\textwidth}
    \centering
    \caption{Family~1: 3D axis}
    \includegraphics[width=\textwidth]{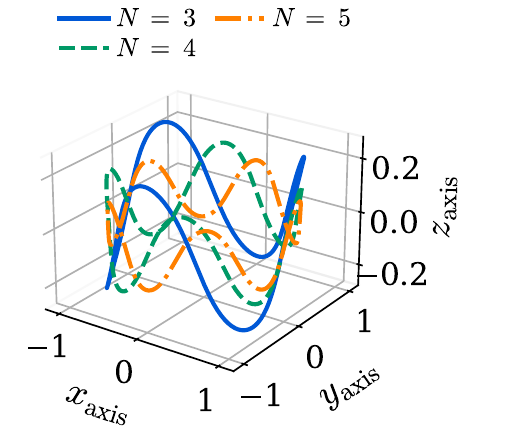}
    \label{fig:3d_axes_all}
  \end{subfigure}
  \hfill
  \begin{subfigure}[t]{0.32\textwidth}
    \centering
    \caption{Family~1: $R_{\mathrm{axis}}(\zeta)$}
    \includegraphics[width=\textwidth]{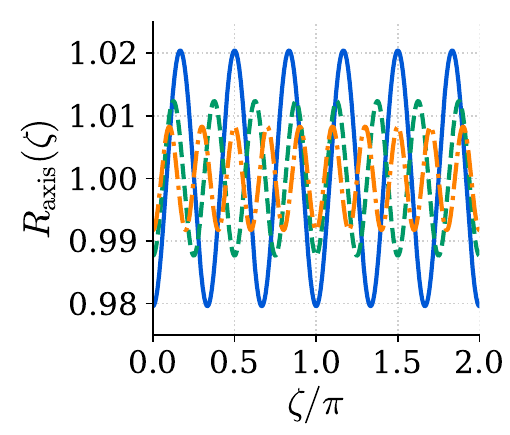}
    \label{fig:axis_xy}
  \end{subfigure}
  \hfill
  \begin{subfigure}[t]{0.32\textwidth}
    \centering
    \caption{Family~1: $z_{\mathrm{axis}}(\zeta)$}
    \includegraphics[width=\textwidth]{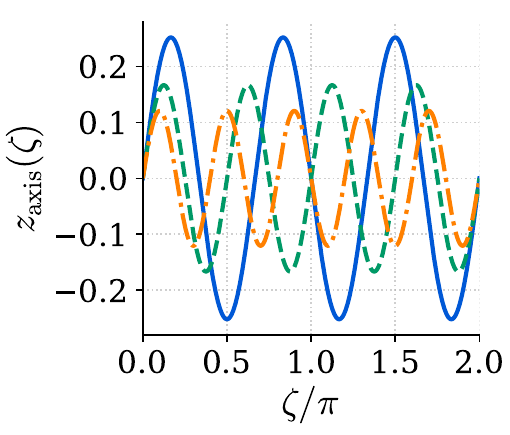}
    \label{fig:axis_z}
  \end{subfigure}
  \\[0.6em]
  \begin{subfigure}[t]{0.32\textwidth}
    \centering
    \caption{Family~2: 3D axis}
    \includegraphics[width=\textwidth]{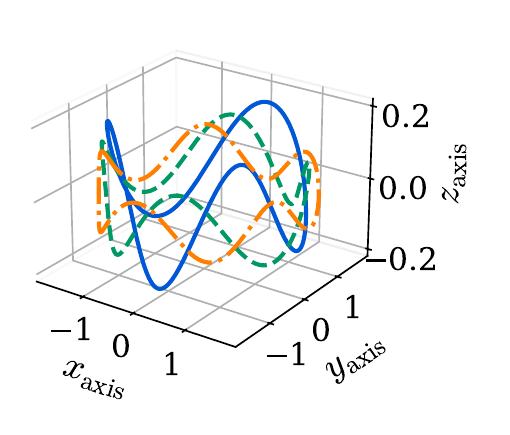}
    \label{fig:3d_axes_fam2}
  \end{subfigure}
  \hfill
  \begin{subfigure}[t]{0.32\textwidth}
    \centering
    \caption{Family~2: $R_{\mathrm{axis}}(\zeta)$}
    \includegraphics[width=\textwidth]{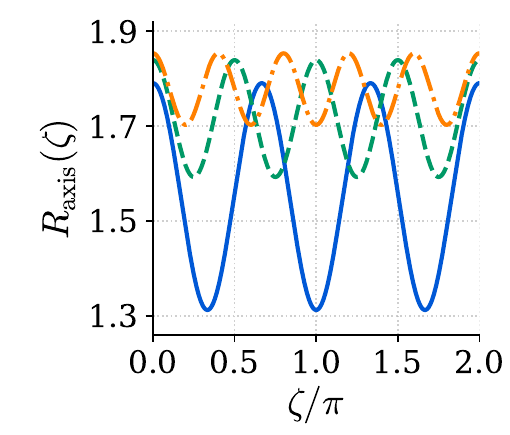}
    \label{fig:axis_xy_fam2}
  \end{subfigure}
  \hfill
  \begin{subfigure}[t]{0.32\textwidth}
    \centering
    \caption{Family~2: $z_{\mathrm{axis}}(\zeta)$}
    \includegraphics[width=\textwidth]{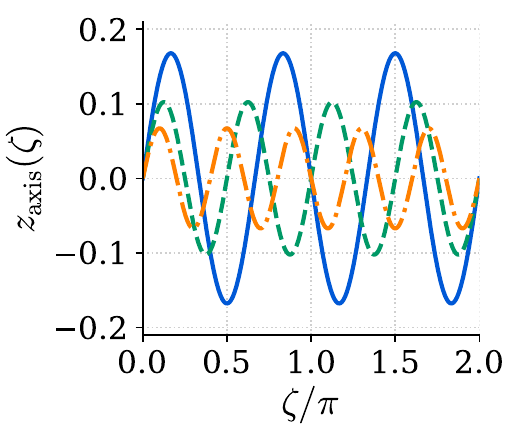}
    \label{fig:axis_z_fam2}
  \end{subfigure}
  \\[0.6em]
  \begin{subfigure}[t]{0.32\textwidth}
    \centering
    \caption{Family~3: 3D axis}
    \includegraphics[width=\textwidth]{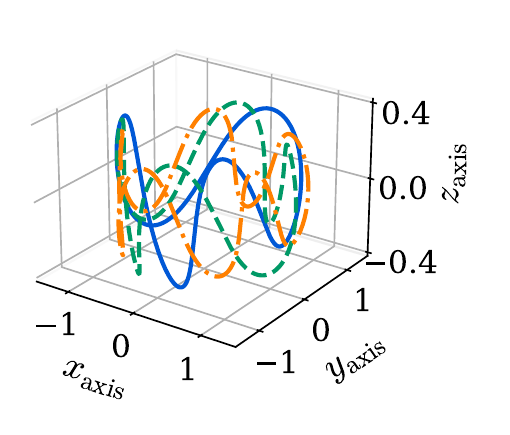}
    \label{fig:3d_axes_fam3}
  \end{subfigure}
  \hfill
  \begin{subfigure}[t]{0.32\textwidth}
    \centering
    \caption{Family~3: $R_{\mathrm{axis}}(\zeta)$}
    \includegraphics[width=\textwidth]{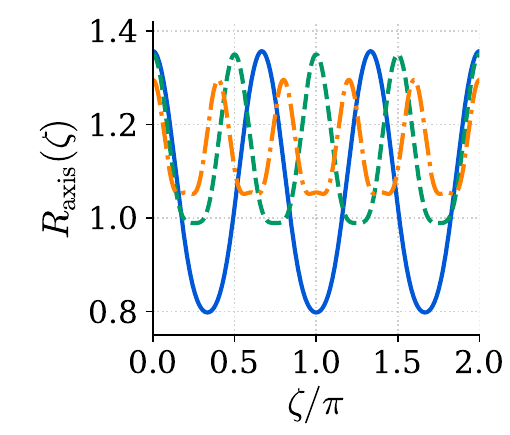}
    \label{fig:axis_xy_fam3}
  \end{subfigure}
  \hfill
  \begin{subfigure}[t]{0.32\textwidth}
    \centering
    \caption{Family~3: $z_{\mathrm{axis}}(\zeta)$}
    \includegraphics[width=\textwidth]{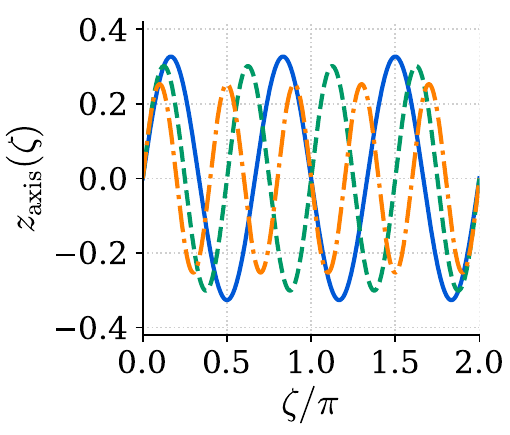}
    \label{fig:axis_z_fam3}
  \end{subfigure}
  \caption{Three-dimensional magnetic axis curves $\xvec_{\mathrm{axis}}(\zeta)$, radial profiles $R_{\mathrm{axis}}(\zeta)$, and vertical profiles $z_{\mathrm{axis}}(\zeta)$ across $N$ for (a)--(c)~Family~1 ($\rho_a = \rho_{0}$), (d)--(f)~Family~2 ($\iota = N$, $\rho_a = 1.12, 1.06, 1.04$), and (g)--(i)~Family~3 ($\iota = N$, $\rho_a = 0.97, 0.89, 0.88$).}
  \label{fig:axis_geometry}
\end{figure}

%======================================================================
Evaluating Eq.~\eqref{eq:xyz_embedding} at $(\rho, \eta) = (\rho_a, 0)$ gives the magnetic axis $\xvec_{\mathrm{axis}}(\zeta)$, i.e.
\begin{equation} \label{eq:axis_general}
\begin{aligned}
  x_{\mathrm{axis}}(\zeta) + i y_{\mathrm{axis}}(\zeta) &= C_1(\rho_a)\,e^{i m_1\zeta} + C_2(\rho_a)\,e^{i m_2\zeta} - C_3(\rho_a)\,e^{i m_3\zeta}, \\
  z_{\mathrm{axis}}(\zeta) &= C_5(\rho_a)\sin(N\zeta).
\end{aligned}
\end{equation}
The squared magnetic axis cylindrical radius is
\begin{equation} \label{eq:R_axis_general}
\begin{aligned}
  R_{\mathrm{axis}}(\zeta)^2 &\coloneq |x_{\mathrm{axis}}(\zeta) + i y_{\mathrm{axis}}(\zeta)|^2 = C_1(\rho_a)^2 + C_2(\rho_a)^2 + C_3(\rho_a)^2 \\
  &\quad + 2 C_2(\rho_a)\big(C_1(\rho_a) - C_3(\rho_a)\big)\cos(N\zeta) - 2 C_1(\rho_a)C_3(\rho_a)\cos(2N\zeta).
\end{aligned}
\end{equation}
In Eq.~\eqref{eq:axis_general}, the $m_j = 1$ harmonic represents the axisymmetric part, while the other two harmonics produce the $\cos(N\zeta)$ and $\cos(2N\zeta)$ radial modulations in Eq.~\eqref{eq:R_axis_general}.
In Family~1 ($m_1 = 1$) with $\rho_a = \rho_0$, $C_2(\rho_a) = 0$ eliminates the $\cos(N\zeta)$ term in Eq.~\eqref{eq:R_axis_general}, resulting in
\begin{equation*}
  R_{\mathrm{axis}}(\zeta) = \rho_0\sqrt{1 - \frac{2\cos(2N\zeta)}{(2N+1)^2} + \frac{1}{(2N+1)^4}}, \qquad
  z_{\mathrm{axis}}(\zeta) = \frac{2\rho_0}{N\sqrt{2N+1}}\sin(N\zeta).
\end{equation*}
As shown in Figures~\ref{fig:3d_axes_all}--\ref{fig:axis_z}, $R_{\mathrm{axis}}(\zeta)$ has $2N$ small oscillations around $\rho_0 = 1$ while $z_{\mathrm{axis}}(\zeta)$ has $N$ vertical oscillations.
By contrast, in Families~2 and~3 (as well as Family~1 with $\rho_a > \rho_0$), $C_2(\rho_a) \neq 0$ retains the $\cos(N\zeta)$ term in Eq.~\eqref{eq:R_axis_general}, resulting in much larger radial oscillations (Figures~\ref{fig:3d_axes_fam2}--\ref{fig:axis_z_fam3}).
Across all three families of equilibria in Figure~\ref{fig:axis_geometry}, both the radial and vertical oscillation amplitudes decrease monotonically with increasing $N$.

%======================================================================
\subsection{Toroidal volume and volume-averaged \texorpdfstring{$\beta_V$}{beta\_V}} \label{subsec:vol_beta}
\begin{figure}[htb!]
  \centering
  \begin{subfigure}[t]{0.48\textwidth}
    \centering
    \caption{Enclosed plasma volume $V(\delta)$}
    \includegraphics[width=\textwidth]{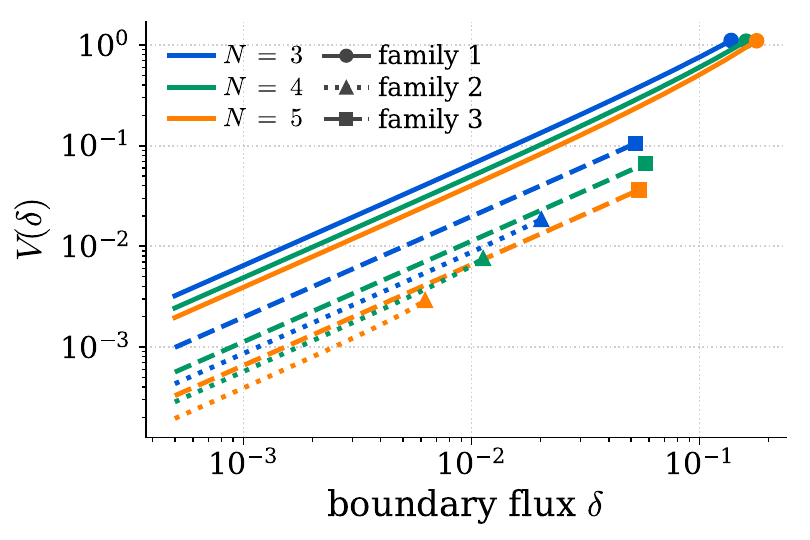}
    \label{fig:volume_delta}
  \end{subfigure}
  \hfill
  \begin{subfigure}[t]{0.48\textwidth}
    \centering
    \caption{Volume-averaged beta $\beta_V(\delta)$}
    \includegraphics[width=\textwidth]{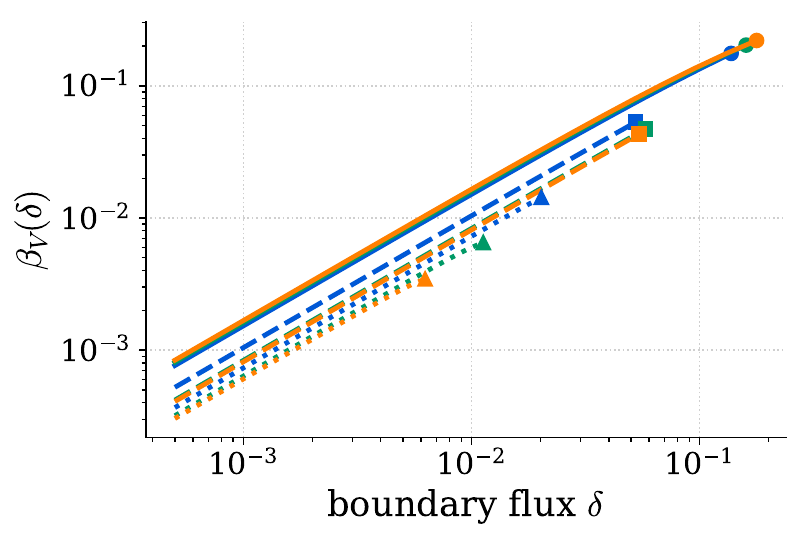}
    \label{fig:beta_delta}
  \end{subfigure}
  \caption{Enclosed plasma volume $V(\delta)$ and volume-averaged beta $\beta_V(\delta)$ versus boundary flux $\delta$ across $N \in \{3, 4, 5\}$ for Family~1 with $\iota = 0$ and $\rho_a = \rho_0$, Family~2 with $\iota = \pm N$, and Family~3 with $\iota = \pm N$. Markers indicate the maximum boundary flux.}
  \label{fig:vol_beta}
\end{figure}

%======================================================================
The volume $V(\delta)$ enclosed by $\psi = \delta$ and the volume-averaged plasma beta $\beta_V(\delta)$ are
\begin{align*}
  V(\delta) &= \int_0^\delta V'(\psi)\,\mathrm{d}\psi, &
  \beta_V(\delta) &\coloneq \frac{2\langle p \rangle_V}{\langle |\Bvec|^2 \rangle_V} = \frac{2 \int_0^\delta \big(p(\delta) + 2(\delta - \psi)\big) V'(\psi)\,\mathrm{d}\psi}{\int_0^\delta (H(\rho_{a}, 0) + 2\psi) V'(\psi)\,\mathrm{d}\psi}, \\
  V'(\psi) &\coloneq 2\pi \int_0^{2\pi} \mathcal{J}_{\psi\theta\zeta}(\psi, \theta)\,\mathrm{d}\theta, &
  \mathcal{J}_{\psi\theta\zeta}(\psi, \theta) &\coloneq \frac{|\mathcal{J}(\rho, \eta)|}{|m_2|\,|\mathrm{d}r/\mathrm{d}\rho|\,|C_4(\rho)|},
\end{align*}
where the volume average of a quantity $f$ is
\begin{equation*}
  \langle f \rangle_V \coloneq \frac{1}{V(\delta)}\int_{0 \le \psi \le \delta} f\,\mathrm{d}V.
\end{equation*}
Figure~\ref{fig:vol_beta} shows the enclosed volume $V(\delta)$ and volume-averaged plasma beta $\beta_V(\delta)$ versus the boundary flux $\delta$ across all three families of equilibria for $N \in \{3, 4, 5\}$.
For Families~2 and~3, we use the same $\rho_a$ values as in Figure~\ref{fig:axis_geometry}, which approximately maximize the peak $\beta_V$ for each $N$.
In all cases, both quantities grow nearly linearly with $\delta$, and $V(\delta)$ decreases with $N$ at fixed $\delta$.
Because the $\iota = \pm N$ branch of Family~2 has thin cross-sections (Figure~\ref{fig:rz_fam2_pos_n3}) and breaks down at a smaller boundary flux $\delta$, it is limited to a small plasma volume and $\beta_V$.
By contrast, Family~3 sustains nested flux surfaces up to a larger boundary flux $\delta$, reaching a larger maximum plasma volume and $\beta_V$ at $\iota = \pm N$.

%======================================================================
\subsection{Ideal-MHD interchange stability} \label{subsec:stability}
%======================================================================
Because every field line closes after one toroidal transit with $\oint |\Bvec|^{-1}\,\mathrm{d}\ell = 2\pi$, the enclosed toroidal magnetic flux is $\Phi_t(\psi) = V(\psi)/2\pi$ and $\mathrm{d}^2 V/\mathrm{d}\Phi_t^2 = 0$, meaning there is neither magnetic shear nor a stabilizing magnetic well ($\mathrm{d}^2 V/\mathrm{d}\Phi_t^2 < 0$) or magnetic hill ($\mathrm{d}^2 V/\mathrm{d}\Phi_t^2 > 0$).
To determine stability, we therefore evaluate the ideal-MHD energy principle~\cite{Bernstein1958, Freidberg2014} directly, under which an equilibrium is unstable if $\delta W(\vec{\xi}) < 0$ for any smooth displacement $\vec{\xi}$ vanishing on the plasma boundary $\psi = \delta$:
\begin{equation} \label{eq:deltaW}
  \delta W(\vec{\xi}) = \frac{1}{2}\int_{0 \le \psi \le \delta} \Big[ |\vec{Q}|^2 + \Jvec \cdot (\vec{\xi} \times \vec{Q}) + (\vec{\xi} \cdot \grad p)(\divg\vec{\xi}) + \gamma p\,(\divg\vec{\xi})^2 \Big]\,\mathrm{d}V,
\end{equation}
where $\vec{Q} \coloneq \curl(\vec{\xi} \times \Bvec)$ is the magnetic field perturbation and $\gamma > 0$ is the adiabatic index.

Setting $\chi(\rho, \eta) \coloneq (\delta - \psi)^3\,\eta$, we consider a trial displacement $\vec{\xi} = \vec{\xi}_0 + \epsilon\,\vec{\xi}_1$~\cite{Spies1974} that vanishes on $\psi = \delta$, with constant $\epsilon$ and
\begin{equation*}
  \vec{\xi}_0 = \frac{1}{\mathcal{J}}\left(\frac{\partial\chi}{\partial\eta}\frac{\partial\xvec}{\partial\rho} - \frac{\partial\chi}{\partial\rho}\frac{\partial\xvec}{\partial\eta}\right), \qquad
  \vec{\xi}_1 = \frac{\chi}{\mathcal{J}}\left(\frac{\partial p}{\partial\eta}\frac{\partial\xvec}{\partial\rho} - \frac{\partial p}{\partial\rho}\frac{\partial\xvec}{\partial\eta}\right).
\end{equation*}
Because $\mathcal{J}(\rho, \eta)$ is independent of $\zeta$ and $\Bvec = \partial_\zeta\xvec = \mathcal{J}(\grad\eta \times \grad\rho)$, the pure interchange $\vec{\xi}_0$ satisfies $\vec{\xi}_0 \times \Bvec = \grad\chi$; it therefore neither perturbs the magnetic field nor compresses the plasma, resulting in $\delta W(\vec{\xi}_0) = 0$.
Adding the flux-surface-tangent displacement $\epsilon\,\vec{\xi}_1$ introduces parallel magnetic and plasma compression governed by
\begin{equation*}
  g(\rho, \eta) \coloneq \frac{2(\delta - \psi)^3}{\mathcal{J}}\frac{\partial\psi}{\partial\rho},
\end{equation*}
such that $\vec{Q} = -\epsilon g\,\Bvec$, $\divg\vec{\xi} = \epsilon g$, and $\vec{\xi} \cdot \grad p = -g$.
Using $\Jvec \cdot (\vec{\xi} \times \vec{Q}) = \epsilon g\,\vec{\xi} \cdot (\Jvec \times \Bvec) = -\epsilon g^2$, Eq.~\eqref{eq:deltaW} simplifies to
\begin{equation*}
  \delta W(\vec{\xi}_0 + \epsilon\,\vec{\xi}_1) = \frac{1}{2}\int_{0 \le \psi \le \delta} g^2 \Big[ -2\epsilon + \epsilon^2\big(|\Bvec|^2 + \gamma p\big) \Big]\,\mathrm{d}V,
\end{equation*}
which is negative for sufficiently small $\epsilon > 0$, proving that all three families of equilibria are ideal-MHD interchange unstable.
%=============================================================================
\section{Exterior vacuum field} \label{sec:vacuum}
%=============================================================================
To benchmark free-boundary equilibrium codes (e.g., VMEC~\cite{hirshman1986three} and DESC~\cite{conlin2024highorderfreeboundary}), we construct a vacuum magnetic field $\Bvec_{\mathrm{vac}}$ in a thin layer outside the plasma boundary $\psi = \delta$ that matches the equilibrium magnetic field $\Bvec$ across the interface:
\begin{equation} \label{eq:vacuum_bvp}
  \curl\Bvec_{\mathrm{vac}} = \vec{0} \quad \text{and} \quad \divg\Bvec_{\mathrm{vac}} = 0 \quad \text{for } \delta < \psi \le \delta_{\mathrm{vac}},
  \qquad \Bvec_{\mathrm{vac}}\big|_{\psi = \delta} = \Bvec\big|_{\psi = \delta},
\end{equation}
where $\delta_{\mathrm{vac}} > \delta$ is close enough to $\delta$ that the coordinate map~\eqref{eq:xyz_embedding} remains invertible.
Matching all components of $\Bvec$ implies no surface current, so pressure balance requires $p(\delta) = 0$, i.e., $p_0 = 2\delta$.

For Family~1 with $\rho_a = \rho_0$, expressing $\Bvec = \partial_\zeta\xvec$ in the reciprocal basis $(\grad\rho, \grad\zeta, \grad\eta)$ yields
\begin{equation} \label{eq:B_scalar_form}
  \Bvec = H(\rho, \eta)\grad\zeta + \grad\Phi(\rho, \zeta, \eta),
\end{equation}
where $\Phi(\rho, \zeta, \eta)$ is the closed-form, single-valued scalar potential
\begin{equation} \label{eq:Phi_def}
  \Phi(\rho, \zeta, \eta) \coloneq \frac{2(N + 1)}{N}\big(C_1(\rho) + (2N + 1)C_3(\rho)\big)\eta\,C_4(\rho)\cos(N\zeta).
\end{equation}

The exterior vacuum magnetic field is
\begin{equation} \label{eq:Bvac_series}
  \Bvec_{\mathrm{vac}} = \big(H(\rho_a, 0) + 2\delta\big)\grad\zeta + \grad\Phi(\rho, \zeta, \eta) + \grad\!\left[\sum_{n=2}^{\infty} \alpha_n(\rho, \zeta, \eta)\big(\psi(\rho, \eta) - \delta\big)^n\right],
\end{equation}
where the coefficients $\alpha_n(\rho, \zeta, \eta)$ are determined purely by differentiation via the recurrence
\begin{equation} \label{eq:alpha_recurrence}
\begin{aligned}
  \alpha_2 &= \frac{\grad\psi \cdot \grad\zeta}{|\grad\psi|^2}, \qquad
  \alpha_3 = \frac{\nabla^2\zeta - 2\,\grad\psi \cdot \grad\alpha_2 - (\nabla^2\psi)\,\alpha_2}{3\,|\grad\psi|^2}, \\[1.0ex]
  \alpha_n &= -\frac{2\,\grad\psi \cdot \grad\alpha_{n-1} + (\nabla^2\psi)\,\alpha_{n-1} + \frac{1}{n-1}\nabla^2\alpha_{n-2}}{n\,|\grad\psi|^2} \qquad \text{for } n \ge 4.
\end{aligned}
\end{equation}

\begin{figure}[htbp]
  \centering
  \begin{subfigure}[t]{0.32\textwidth}
    \centering
    \caption{Scaling with $\psi - \delta$}
    \includegraphics[width=\textwidth]{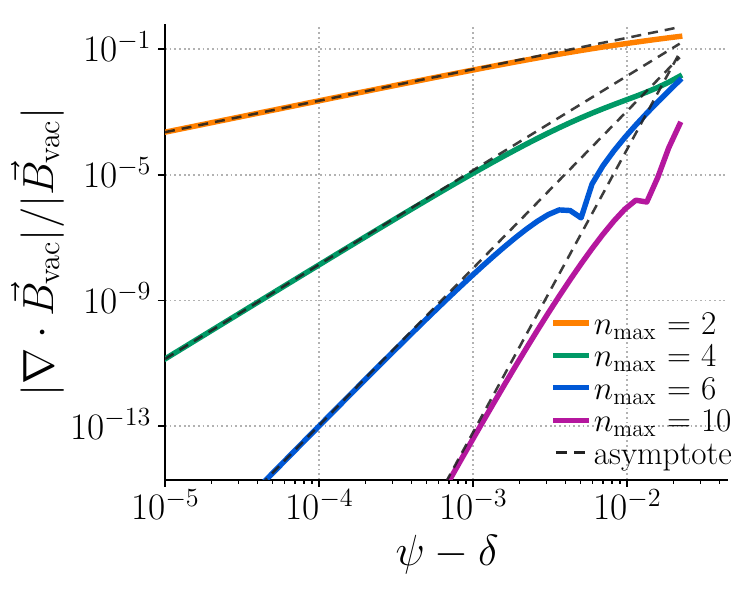}
    \label{fig:vacuum_scaling}
  \end{subfigure}
  \hfill
  \begin{subfigure}[t]{0.32\textwidth}
    \centering
    \caption{$\divg\Bvec_{\mathrm{vac}}$ residual at $\phi = 0$}
    \includegraphics[width=\textwidth]{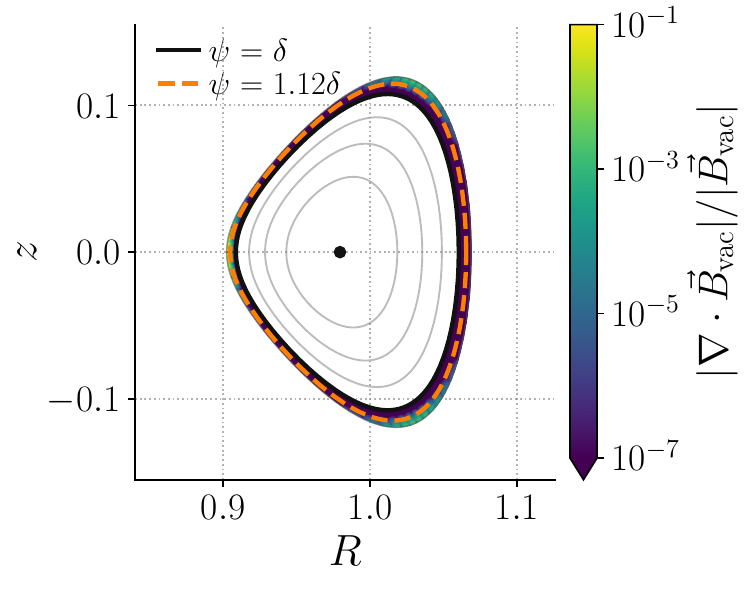}
    \label{fig:vacuum_rz}
  \end{subfigure}
  \hfill
  \begin{subfigure}[t]{0.32\textwidth}
    \centering
    \caption{$\Bvec$ and $\Bvec_{\mathrm{vac}}$ at $\phi = 0$}
    \includegraphics[width=\textwidth]{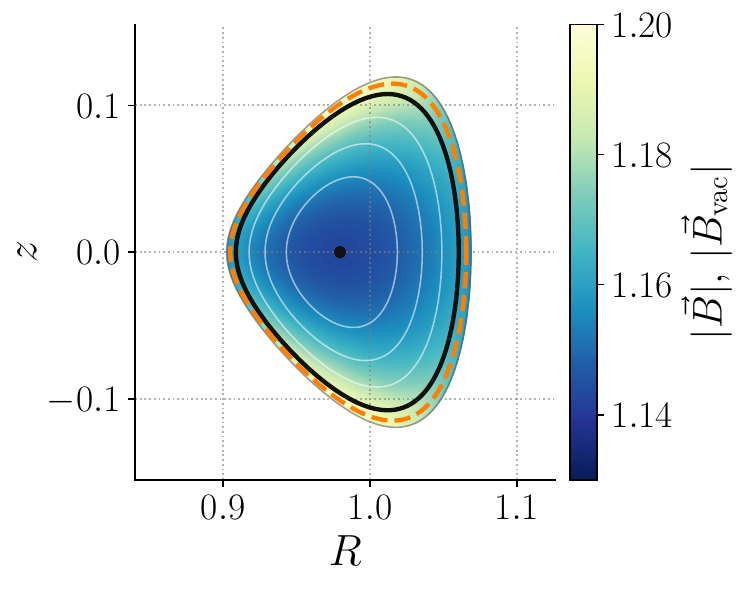}
    \label{fig:vacuum_bmag}
  \end{subfigure}
  \caption{Verification of the truncated exterior vacuum field in Eq.~\eqref{eq:Bvac_series} for Family~1 with $N = 3$, $\rho_a = \rho_0$, and plasma boundary $\delta = 0.02$. (a)~Normalized divergence residual along $\theta = \pi/4$ and $\zeta = \pi/4N$ versus distance $\psi - \delta$ from the boundary across truncation orders $n_{\max}$. (b)~Poloidal cross-section at $\phi = 0$ showing the normalized divergence residual for $n_{\max} = 10$. (c)~Poloidal cross-section at $\phi = 0$ showing the magnetic field magnitude $|\Bvec|$ in the interior plasma ($0 \le \psi \le \delta$) and $|\Bvec_{\mathrm{vac}}|$ in the exterior shell ($\delta \le \psi \le \delta_{\mathrm{vac}} = 1.2\delta$) for $n_{\max} = 10$.}
  \label{fig:vacuum}
\end{figure}

To verify that Eqs.~\eqref{eq:Bvac_series}--\eqref{eq:alpha_recurrence} satisfy Eq.~\eqref{eq:vacuum_bvp}, note first that $\curl\Bvec_{\mathrm{vac}} = \vec{0}$ holds identically because every term in Eq.~\eqref{eq:Bvac_series} is locally a gradient.
Second, subtracting Eq.~\eqref{eq:B_scalar_form} from Eq.~\eqref{eq:Bvac_series} gives
\begin{equation} \label{eq:Bvac_minus_B}
  \Bvec_{\mathrm{vac}} - \Bvec = -2\big(\psi(\rho, \eta) - \delta\big)\grad\zeta + \grad\!\left[\sum_{n=2}^{\infty} \alpha_n(\rho, \zeta, \eta)\big(\psi(\rho, \eta) - \delta\big)^n\right],
\end{equation}
where all terms vanish on $\psi = \delta$, confirming the boundary condition in Eq.~\eqref{eq:vacuum_bvp}.
Third, taking the divergence of Eq.~\eqref{eq:Bvac_minus_B} and using $\divg\Bvec = 0$ from Eq.~\eqref{eq:jacobian_exact} yields
\begin{equation*}
  \divg\Bvec_{\mathrm{vac}} = -2\,\grad\psi \cdot \grad\zeta - 2\big(\psi - \delta\big)\nabla^2\zeta + \nabla^2\!\left[\sum_{n=2}^{\infty} \alpha_n\big(\psi - \delta\big)^n\right],
\end{equation*}
whose coefficients of each power of $\psi - \delta$ vanish by the recurrence relation in Eq.~\eqref{eq:alpha_recurrence}.
Truncating the series at $n = n_{\max}$ preserves $\curl\Bvec_{\mathrm{vac}} = \vec{0}$ and $\Bvec_{\mathrm{vac}} = \Bvec$ on $\psi = \delta$ exactly, while leaving $\divg\Bvec_{\mathrm{vac}} = \mathcal{O}\big((\psi - \delta)^{n_{\max}-1}\big)$, as verified in Figure~\ref{fig:vacuum_scaling}.
For $N = 3$, $\delta = 0.02$, and $n_{\max} = 10$, Figures~\ref{fig:vacuum_rz} and~\ref{fig:vacuum_bmag} show the divergence residual $|\divg\Bvec_{\mathrm{vac}}| / |\Bvec_{\mathrm{vac}}|$ and the continuous magnetic field strength across $\psi = \delta$ in the $\phi = 0$ cross-section for $\delta_{\mathrm{vac}} = 1.2\delta$.

Note that for $\psi > \delta$, $\psi(\rho, \eta)$ serves only as a spatial coordinate rather than a flux function of $\Bvec_{\mathrm{vac}}$, since $\Bvec_{\mathrm{vac}} \cdot \grad\psi = \mathcal{O}\big((\psi - \delta)^2\big) \neq 0$ and exterior field lines cross surfaces of constant $\psi$.
Nevertheless, because $\Bvec_{\mathrm{vac}}$ inherits both discrete $N$-fold rotational symmetry~\eqref{eq:nfp_symmetry} and reflection symmetry~\eqref{eq:reflection_def} across $\phi_* = \pi/2N$, every exterior field line near $\psi = \delta$ closes after one toroidal transit ($\iota = 0$).
Consequently, prescribing $\psi_{\mathrm{vac}} = \psi$ on a poloidal cross-section and integrating $\Bvec_{\mathrm{vac}} \cdot \grad\psi_{\mathrm{vac}} = 0$ along the closed field lines defines a smooth, single-valued function $\psi_{\mathrm{vac}}(\xvec) = \psi(\rho, \eta) + \mathcal{O}\big((\psi - \delta)^2\big)$ whose level sets form nested toroidal magnetic surfaces.
Where the series in Eq.~\eqref{eq:Bvac_series} converges, these level sets are exact flux surfaces, whereas truncating at $n = n_{\max}$ preserves $\Bvec_{\mathrm{vac}} \cdot \grad\psi_{\mathrm{vac}} = 0$ while conserving magnetic flux up to $\mathcal{O}\big((\psi - \delta)^{n_{\max}-1}\big)$.

In free-boundary benchmarks, $\Bvec$ and $\Bvec_{\mathrm{vac}}$ can be used either by prescribing $\Bvec_{\mathrm{vac}} \cdot \hat{n}$ on a control surface $\delta < \psi < \delta_{\mathrm{vac}}$ (e.g.\ $\psi = 1.12\delta$ in Figures~\ref{fig:vacuum_rz} and~\ref{fig:vacuum_bmag}), or by supplying the external field $\Bvec_{\mathrm{ext}} = \Bvec_{\mathrm{vac}} - \Bvec_{\mathrm{p}}$, where $\Bvec_{\mathrm{p}}$ is the Biot--Savart field of the plasma current $\Jvec = \curl\Bvec = 2\grad\psi\times\grad\zeta$.
For the remaining equilibria with $C_2(\rho) \neq 0$, the plasma carries a net toroidal current, so a matching curl-free field must also include a multivalued poloidal potential. Constructing this potential requires the radial extent of the boundary cross-section, which in general must be found numerically by root finding, so we restrict the vacuum field to Family~1 with $\rho_a = \rho_0$.

%=============================================================================
\section{Discussion} \label{sec:summary}
%=============================================================================
This paper introduces three closed-form families of non-axisymmetric ideal MHD equilibria with nested toroidal flux surfaces, stellarator symmetry, and non-constant pressure for any integer number of field periods $N \ge 1$.
\citet{lortz1970existenz} previously proved the existence of $\iota = 0$ toroidal equilibria in the limit of low plasma beta via an iterative scheme, relying on reflection symmetry.
Family~1 with $\rho_a = \rho_0$ provides an explicit realization of the~\citet{lortz1970existenz} equilibria in elementary functions that guarantees nested toroidal flux surfaces, holds at finite plasma beta for arbitrary $N \ge 1$.
Relaxing reflection symmetry yields additional $\iota = 0$ solutions in Family~1 and Family~2 ($N \ge 2$), as well as equilibria with nonzero rotational transform $\iota = \pm N$ in Family~2 and Family~3 for arbitrary $N \ge 3$.

The present equilibria are the first closed-form solutions to provide benchmarks for verifying three-dimensional MHD equilibrium solvers in the $N \ge 3$ regime.
For Family~1 with $\rho_a = \rho_0$, the matched exterior vacuum field of Section~\ref{sec:vacuum} further extends this benchmark to free-boundary equilibrium codes.
Now that several distinct classes of exact non-axisymmetric toroidal equilibria have been found, there are likely many more waiting to be discovered.
It would be interesting to investigate whether a more general mathematical theory governs and unifies these classes of exact three-dimensional solutions.

%=================================
\section{Acknowledgments on AI assistance} \label{sec:AI_agents}
%=================================
This research was conducted with the assistance of Gemini 4 multi-agent workflows within the Google Antigravity environment. The authors posed the problem of discovering toroidal equilibria for $N \ge 3$ and guided the agents through the analysis. Gemini also assisted in preparing the manuscript and the accompanying \texttt{SymPy} verification and plotting scripts. All mathematical derivations and symbolic checks were reviewed and validated by the authors, who assume full responsibility for the manuscript.

%=============================================================================
\appendix
\section{Origin of the equilibria} \label{app:derivation}
%=============================================================================
Exploiting symmetries to restrict the form of a solution ansatz is a standard route to exact solutions of differential equations~\cite{Olver1993, Hydon2000}, including closed-form MHD equilibria~\cite{KellerCheviakov2023}, and symbolic packages are routinely used to automate the resulting calculations~\cite{DimasTsoubelis2005, Meurer2017, Oliveri2021}. In that spirit, we outline how the equilibria in Section~\ref{sec:equilibrium} were found by perturbing the axisymmetric coordinate map and imposing discrete $N$-fold rotational symmetry~\eqref{eq:nfp_symmetry} to derive Eqs.~\eqref{eq:xyz_embedding}--\eqref{eq:psi_p} from the ideal MHD system~\eqref{eq:mhd_system}.

%----------------------------------------------------------------------
\subsection{Covariant field-line formulation} \label{app:subsec_covariant}
%----------------------------------------------------------------------
We formulate the problem as an inverse coordinate map $\xvec(\rho, \zeta, \eta)$ with Jacobian determinant 
\begin{equation*}
    \mathcal{J} = \det\big(\partial\xvec/\partial(\rho, \zeta, \eta)\big).
\end{equation*}
Aligning the magnetic field with the $\zeta$-coordinate curves via $\Bvec = \partial_\zeta\xvec$, the directional derivative along $\Bvec$ is $\Bvec \cdot \grad = \partial_\zeta$, and the divergence-free condition reduces to
\begin{equation} \label{eq:app_divB}
  \divg\Bvec = \frac{1}{\mathcal{J}}\frac{\partial\mathcal{J}}{\partial\zeta} = 0 \quad \Longleftrightarrow \quad \frac{\partial\mathcal{J}}{\partial\zeta} = 0.
\end{equation}
We first rewrite $\Jvec \times \Bvec$ using the vector identity
\begin{equation*}
  \Jvec \times \Bvec = (\Bvec\cdot\grad)\Bvec - \frac{1}{2}\grad|\Bvec|^2 = \frac{\partial^2\xvec}{\partial\zeta^2} - \frac{1}{2}\grad\!\left(\left|\frac{\partial\xvec}{\partial\zeta}\right|^2\right).
\end{equation*}
To express the force balance equation $\Jvec \times \Bvec = \grad p$ in the field-aligned coordinates, we take its dot product with each of the tangent vectors $\frac{\partial\xvec}{\partial\rho}$, $\frac{\partial\xvec}{\partial\eta}$, and $\frac{\partial\xvec}{\partial\zeta}$, which yields
\begin{align}
 \frac{\partial\xvec}{\partial\rho}\cdot(\Jvec\times\Bvec) &= \frac{\partial g_{\rho\zeta}}{\partial\zeta} - \frac{\partial g_{\zeta\zeta}}{\partial\rho} = \frac{\partial p}{\partial\rho}, \label{eq:rho-component}\\
 \frac{\partial\xvec}{\partial\eta}\cdot(\Jvec\times\Bvec) &= \frac{\partial g_{\eta\zeta}}{\partial\zeta} - \frac{\partial g_{\zeta\zeta}}{\partial\eta} = \frac{\partial p}{\partial\eta}, \label{eq:eta-component} \\
 \frac{\partial\xvec}{\partial\zeta}\cdot(\Jvec\times\Bvec) &= \frac{\partial g_{\zeta\zeta}}{\partial\zeta} - \frac{\partial g_{\zeta\zeta}}{\partial\zeta} = 0 = \frac{\partial p}{\partial\zeta}, \label{eq:zeta-component}
\end{align}
where the metric components are
\begin{equation}\label{eq:g_def}
  g_{\rho\zeta} = \frac{\partial\xvec}{\partial\rho} \cdot \frac{\partial\xvec}{\partial\zeta}, \qquad
  g_{\eta\zeta} = \frac{\partial\xvec}{\partial\eta} \cdot \frac{\partial\xvec}{\partial\zeta}, \qquad
  g_{\zeta\zeta} = \left|\frac{\partial\xvec}{\partial\zeta}\right|^2 = |\Bvec|^2.
\end{equation}
For a $\zeta$-independent pressure $p(\rho, \eta)$, Eq.~\eqref{eq:zeta-component} is satisfied automatically, reducing the three-dimensional MHD system~\eqref{eq:mhd_system} to Eqs.~\eqref{eq:app_divB}--\eqref{eq:eta-component}.

%----------------------------------------------------------------------
\subsection{Coordinate ansatz} \label{app:subsec_ansatz}
%----------------------------------------------------------------------
Starting from the axisymmetric coordinate mapping $x + iy = \rho\,e^{i\zeta}$, $z = \eta$, we add an $N$-periodic vertical oscillation
\begin{equation*} 
  z(\rho, \zeta, \eta) = \eta + C_5(\rho, \eta)\sin(N\zeta),
\end{equation*}
resulting in the metric~\eqref{eq:g_def} components
\begin{align*}
  g_{\rho\zeta} &= \frac{\partial x}{\partial\rho}\frac{\partial x}{\partial\zeta} + \frac{\partial y}{\partial\rho}\frac{\partial y}{\partial\zeta} + \frac{N C_5}{2}\frac{\partial C_5}{\partial\rho}\sin(2N\zeta), \\
  g_{\eta\zeta} &= \frac{\partial x}{\partial\eta}\frac{\partial x}{\partial\zeta} + \frac{\partial y}{\partial\eta}\frac{\partial y}{\partial\zeta} + N C_5\cos(N\zeta) + \frac{N C_5}{2}\frac{\partial C_5}{\partial\eta}\sin(2N\zeta), \\
  g_{\zeta\zeta} &= \left(\frac{\partial x}{\partial\zeta}\right)^2 + \left(\frac{\partial y}{\partial\zeta}\right)^2 + \frac{N^2 C_5^2}{2}\big(1 + \cos(2N\zeta)\big).
\end{align*}
Since $p(\rho, \eta)$ is independent of $\zeta$, the resulting harmonics $N$ and $2N$ in Eqs.~\eqref{eq:rho-component}--\eqref{eq:eta-component} must be cancelled by matching harmonics in $x + iy$.

The $N$-fold symmetry~\eqref{eq:nfp_symmetry} restricts $x + iy$ to the Fourier series
\begin{equation*}
  x + iy = e^{i\zeta}\sum_{n} F_n(\rho, \eta)\,e^{i nN\zeta} = \sum_{n} F_n(\rho, \eta)\,e^{i(1 + nN)\zeta}.
\end{equation*}
The only set of modes producing beat frequencies $N$ and $2N$ without higher ones is three consecutive harmonics $n \in \{k-1, k, k+1\}$ centered at an integer index $k$.
For the magnetic axis to wind once around the $z$-axis, the fundamental mode $n = 0$ must be one of these three harmonics.
Choosing $n = 0$ to be the first, second, or third harmonic gives $k = 1$, $0$, or $-1$, which defines the three families of equilibria with mode numbers in Eq.~\eqref{eq:mode_defs}.

Cancelling the $2N\zeta$ harmonics in Eqs.~\eqref{eq:app_divB} and~\eqref{eq:eta-component} allows $F_{k \pm 1}$ and $C_5$ to be independent of $\eta$, so only $F_k$ depends on $\eta$.
Taking $F_k$ to be linear in $\eta$, we parameterize the amplitudes in terms of real functions $C_1(\rho), \dots, C_5(\rho)$ as
\begin{equation*}
  F_{k-1}(\rho) = C_1(\rho), \qquad
  F_k(\rho, \eta) = C_2(\rho) + i\eta\,C_4(\rho), \qquad
  F_{k+1}(\rho) = -C_3(\rho),
\end{equation*}
which yields the coordinate ansatz~\eqref{eq:xyz_embedding}:
\begin{equation}\label{eq:coordinate_anstaz}
\begin{aligned}
  x + iy &= C_1(\rho)\,e^{i m_1\zeta} + \big(C_2(\rho) + i\eta\,C_4(\rho)\big)e^{i m_2\zeta} - C_3(\rho)\,e^{i m_3\zeta}, \\
  z &= \eta + C_5(\rho)\sin(N\zeta).
\end{aligned}
\end{equation}

%----------------------------------------------------------------------
\subsection{Determining the auxiliary functions \texorpdfstring{$C_1, \dots, C_5$}{C\_1, ..., C\_5}} \label{app:subsec_harmonic_solve}
%----------------------------------------------------------------------
Substituting the ansatz~\eqref{eq:coordinate_anstaz} into Eqs.~\eqref{eq:app_divB}--\eqref{eq:eta-component}, the coefficients of each $\zeta$-harmonic must vanish independently, while the $\zeta$-independent terms determine the pressure $p(\rho, \eta)$.
We solve these conditions in three steps:

\paragraph{1. Second harmonics ($2N\zeta$).}
Setting the $2N\zeta$ coefficients in Eqs.~\eqref{eq:app_divB}--\eqref{eq:eta-component} to zero yields the Euler differential equations
\begin{equation*}
  \rho\frac{\mathrm{d}C_1}{\mathrm{d}\rho} - m_1 C_1(\rho) = 0, \qquad
  \rho\frac{\mathrm{d}C_3}{\mathrm{d}\rho} + m_3 C_3(\rho) = 0, \qquad
  \rho\frac{\mathrm{d}C_5}{\mathrm{d}\rho} + N C_5(\rho) = 0,
\end{equation*}
after choosing the radial coordinate $\rho$ to fix the first power law.
Integrating directly gives
\begin{equation} \label{eq:app_step12}
  C_1(\rho) = c_1\,\rho^{m_1}, \qquad
  C_3(\rho) = \frac{c_3}{\rho^{m_3}}, \qquad
  C_5(\rho) = \frac{c_5}{\rho^{N}},
\end{equation}
where $c_1, c_3, c_5$ are real integration constants.

\paragraph{2. Fundamental harmonics ($N\zeta$) in force balance.}
With Eq.~\eqref{eq:app_step12}, setting the $N\zeta$ coefficients in the force-balance equations~\eqref{eq:rho-component}--\eqref{eq:eta-component} to zero yields
\begin{equation*}
  D_+(\rho) C_4(\rho) = N^2 C_5(\rho), \qquad
  \rho\frac{\mathrm{d}}{\mathrm{d}\rho}\big(D_-(\rho) C_2(\rho)\big) + N D_-(\rho) C_2(\rho) = 0,
\end{equation*}
with $D_\pm(\rho)$ defined in Eq.~\eqref{eq:D_C2_defs}.
Solving the algebraic relation for $C_4(\rho)$ and integrating the differential equation from $\rho_a$ to $\rho$ for $C_2(\rho)$ gives
\begin{equation} \label{eq:app_step3}
  C_4(\rho) = \frac{N^2 C_5(\rho)}{D_+(\rho)}, \qquad
  C_2(\rho) = C_2(\rho_a)\left(\frac{\rho_a}{\rho}\right)^{N}\frac{D_-(\rho_a)}{D_-(\rho)}.
\end{equation}

\paragraph{3. Divergence-free closure and $\zeta$-independent terms.}
Substituting Eqs.~\eqref{eq:app_step12}--\eqref{eq:app_step3} into the remaining $N\zeta$ coefficient of Eq.~\eqref{eq:app_divB} gives the consistency condition
\begin{equation} \label{eq:app_closure}
  N^2 c_5^2 = 4 m_1 m_3 c_1 c_3 > 0 \quad \Longleftrightarrow \quad N^2 C_5(\rho)^2 = 4 m_1 m_3 C_1(\rho) C_3(\rho).
\end{equation}
The constants $c_1$ and $c_3$ can be fixed without loss of generality in terms of a single length scale $\rho_0 > 0$, which approximates the major radius:
\begin{equation*}
  c_1 = \frac{\operatorname{sign}(m_1)\,\rho_0^{1 - m_1}}{m_1^2}, \qquad
  c_3 = \frac{\rho_0^{1 + m_3}}{m_3^2}, \qquad
  c_5 = \frac{2\,\rho_0^{1 + N}}{N\sqrt{|m_1|\,m_3}},
\end{equation*}
where $c_5$ follows from Eq.~\eqref{eq:app_closure}, taking the positive root without loss of generality. The constants above recover Eq.~\eqref{eq:C}.
The $\zeta$-independent terms in Eqs.~\eqref{eq:rho-component}--\eqref{eq:eta-component} then reduce to
\begin{equation*}
  \frac{\partial p}{\partial\rho} = -\frac{\partial H}{\partial\rho}, \qquad
  \frac{\partial p}{\partial\eta} = -\frac{\partial H}{\partial\eta},
\end{equation*}
with $H(\rho, \eta)$ defined in Eq.~\eqref{eq:psi_p}, which integrate to Eq.~\eqref{eq:psi_p}.
Setting the radial pressure derivative to zero at the magnetic axis $\rho = \rho_a$ gives Eq.~\eqref{eq:D_C2_defs}.

%=================================
\bibliographystyle{unsrtnat}
\bibliography{reference}
%=================================
\end{document}